\documentclass[pre,twocolumn,floatfix,superscriptaddress]{revtex4}
\usepackage{graphicx,amsfonts,amssymb,amsmath, hyperref,multirow,array}
\usepackage[applemac]{inputenc}

\newif\ifhyper
\hypertrue
\ifhyper
\hypersetup{
  citecolor = {green},
  colorlinks = {true}, 
  urlcolor = {blue} 
}
\fi

\newlength{\ldag}
\begin{document}

\title{Rheology of granular liquids in extensional flows: Beyond the $\mu(\mathcal{I})$-law}

\author{O. Coquand} 
\email{oliver.coquand@dlr.de}
\affiliation{Institut f\"ur Materialphysik im Weltraum, Deutsches Zentrum f\"ur Luft- und Raumfahrt (DLR), 51170 K\"oln, Germany}


\author{M. Sperl} 
\email{matthias.sperl@dlr.de}
\affiliation{Institut f\"ur Materialphysik im Weltraum, Deutsches Zentrum f\"ur Luft- und Raumfahrt (DLR), 51170 Cologne, Germany}
\affiliation{Institut f\"ur Theoretische Physik, Universit\"at zu K\"oln, 50937 Cologne, Germany}

\date{}


\begin{abstract}
	The {\bf G}ranular {\bf I}ntegration {\bf T}hrough {\bf T}ransients (GITT) formalism gives a theoretical description
	of the rheology of moderately dense granular flows and suspensions.
	In this work, we extend the GITT equations beyond the case of simple shear flows studied before.
	Applying this to the particular example of extensional flows, we show that the predicted behavior is somewhat different
	from that of the more frequently studied simple shear case, as illustrated by the possibility of non monotonous evolution of the
	effective friction coefficient $\mu$ with the inertial number $\mathcal{I}$.
	By the reduction of the GITT equations to simple toy models, we provide a generalization of the $\mu(\mathcal{I})$-law true for any type of flow deformation.
	Our analysis also includes a study of the Trouton ratio, which is shown to behave quite similarly to that of dense colloidal suspensions.
\end{abstract}

\maketitle

\section{Introduction}

	Granular matter encompasses all systems whose elementary constituents are large particles (typically bigger than 100 $\mu$m) \cite{Andreotti13}.
	Such particles are therefore quite sensitive to the gravitational field on Earth. As a result, most granular flows we can observe around us are dense flows,
	with a packing fraction $\varphi$ typically bigger than 40\%.
	In those conditions, granular matter is in the so called \textit{granular-liquid} state \cite{Andreotti13}.
	Granular liquids have been at the center of an intense research activity both in fundamental physics \cite{GDR04,DaCruz05,Jop05,Jop06,Pouliquen06,Forterre08,Peyneau08,
	Lagree11,Tankeo13,Clavaud17}, and at the interface between physics and geosciences \cite{Savage79,Savage98,Pouliquen02,Cassar05,Frey10,Gueugneau17,Ogburn17,Salmanidou17,Pahtz20}
	or biology \cite{Forterre18,Berut18,Ruhs20}.

	Granular liquids fall into the category of complex liquids, meaning that their macroscopic behavior is somewhat between that of a solid and that
	of a simple liquid.
	A convenient way to quantify how far from a simple liquid the system behaves is to study its \textit{effective friction coefficient} $\mu$.
	By analogy with the Coulomb law of solid friction, $\mu$ can be used to determine whether a granular-liquid going down a slope with a given angle
	can develop a stationary flow or not \cite{Savage89}.
	Thus a simple liquid is expected to have $\mu=0$, whereas more complex soft materials have higher $\mu$ as their behavior becomes increasingly solid-like.
	Unlike solids however, the effective friction coefficient of complex liquids typically depends on the shear rate in a given flow configuration.

	One of the most remarkable properties of granular liquids is that $\mu$ obeys a universal scaling law as a function of the shear rate $\dot\gamma$, or more
	precisely a dimensionless shear rate, called the inertial number $\mathcal{I}$ \cite{GDR04}.
	This scaling law is called the $\mu(\mathcal{I})$-law, and can be written as:
	\begin{equation}
	\label{eqMuI}
		\mu(\mathcal{I}) = \mu_1 + \frac{\mu_2 - \mu_1}{1 + \mathcal{I}_0/\mathcal{I}}\;,
	\end{equation}
	where $\mu_1$, $\mu_2$ and $\mathcal{I}_0$ are characteristics of the material.
	The combination of its simplicity, universality and ability to provide a satisfactory phenomenological law to describe experimental and numerical data
	\cite{Pouliquen06,Forterre08,Tapia19,Fullard19} makes the $\mu(\mathcal{I})$-law a particularly powerful tool.

	Providing a theoretical framework to account for those phenomena on the other hand is still a challenging task.
	In that respect, one promising candidate is the \textit{Granular Integration Through Transients} formalism (GITT) \cite{Kranz18,Kranz20,Coquand20f,Coquand20g}.
	It has been shown that GITT predictions are quantitatively compatible with the existing experimental and numerical literature \cite{Coquand20f}, and that
	GITT equations can be broken down to analytically tractable toy models from which one can derive the expression Eq.~(\ref{eqMuI}) \cite{Coquand20g}.
	Moreover, the GITT framework can easily be extended to granular suspensions, where it can be used to provide simple phenomenological laws analogous to Eq.~(\ref{eqMuI})
	\cite{Coquand20g} in cases where no consensus exists yet \cite{Boyer11,Pahtz19,Tapia19,Suzuki19}.

	However the equations derived in previous works \cite{Kranz18,Kranz20,Coquand20f,Coquand20g} apply only to simple shear flows.
	While this case is relevant to a number of natural flows, such as avalanches for example \cite{Savage89,Savage98},
	it does not provide a full description of the possible rheological behaviors
	of granular liquids.
	This work presents a generalization of the GITT equations to all steady incompressible flows.
	Through the study of the particular example of extensional flows, we show that the law (\ref{eqMuI}) does not account for all the phenomenology
	of granular rheology when the applied stress is not pure shear.
	Due to the method outlined in \cite{Coquand20g} we propose alternatives to the $\mu(\mathcal{I})$-law for granular liquids and granular suspensions that can be tested
	experimentally or numerically.
	This is all the more important that to the best of our knowledge, the evolution of $\mu$ in granular liquids and suspensions under
	extensional flows we report here is quite new. Indeed, even though a number of recent studies have addressed the rheology of non-Brownian suspensions under
	extensional flows despite experimental difficulties \cite{BischoffWhite10,Dai17,Majumdar17,Chateau18,James18,Tanner18,Tanner18r,Tanner19,Tanner20,Shende21},
	none of them proposed a study of the evolution of the effective friction coefficient in those conditions.

	The paper is organized as follows: we first derive the general GITT equations.
	Then, we derive from them a proper framework to describe the evolution of the observables of granular-liquid rheology.
	Third, the next section presents the application
	of this formalism to the particular case of extensional flows, both planar and uniaxial.
	Finally, we conclude.

\section{General GITT equations for incompressible stationary flows}

	The full derivation of the GITT equations from first principles is quite long, we therefore restrict ourselves in the following to the
	part most relevant to our purpose and refer the interested reader to the detailed work \cite{Kranz20}.

	\subsection{The Integration Through Transients formalism}

		Let us consider a granular liquid consisting of $N$ particles, represented by frictionless hard spheres interacting through
		dissipative collisions with a restitution coefficient $\varepsilon$.
		This is a less idealized model than it may appear at first sight \cite{Voivret09,Pahtz20,Coquand20f}.
		Indeed, interparticle friction becomes relevant only for the rheology of very dense granular liquids close to the density of the granular solid
		\cite{Peyneau08,Clavaud17,Coquand20f}.
		We thus restrict ourselves to high densities but sufficiently far away from the friction dominated regime.
		It should be noted however that defined in that way, the liquid state of granular matter does not extend up to the transition
		to the solid state which depends on the value of the interparticle friction coefficient \cite{Ikeda12,Ikeda13,DeGiuli15,DeGiuli16,DeGiuli17a}.
		The transition between the liquid and interparticle friction dominated regime has been estimated to take place for inertial numbers
		$\mathcal{I}\lesssim 0.003$ \cite{DeGiuli16}.

		The rheology of the granular-liquid under consideration is studied in a stationary flow defined by the average velocity profile
		$\mathbf{v}=\kappa\cdot\mathbf{r}$, $\mathbf{r}$ being the position vector.
		Note that $\kappa$ does not need to be symmetric, such as for the simple shear flow, for example,
		where it can be defined by $\kappa_{ij} = \dot\gamma \delta_{ix}\delta_{jy}$.
		In addition, we restrict ourselves to incompressible flows, so that Tr$(\kappa) = 0$.

		The dynamics of the system is described by a mode coupling equation of motion.
		The full derivation of this equation can be found in \cite{Kranz20}.
		Since it is not modified by the introduction of a more general flow matrix $\kappa$, and does not play a central
		role in what follows, we just recall its most salient features.

		In \textit{Mode Coupling Theory} (MCT), the dynamical evolution of the system is studied by use of its dynamical structure
		factor $\Phi_q(t)$ which is the normalized density-density correlation function: $\Phi_q(t) = \left<\rho_q(t)\rho_{-q}\right>/S_q$,
		where $S_q = \left<\rho_q\rho_{-q}\right>$ is the static structure factor \cite{Goetze08,Fuchs02,Fuchs03,Fuchs09,Brader09,Kranz13,Kranz18,Kranz20}.
		The general MCT equation has the structure of a Mori-Zwanzig equation:
		\begin{equation}
		\label{eqMCT}
			\begin{split}
				\ddot{\Phi}_{q}(t) &+ \nu_{q}\dot{\Phi}_{q}(t)+ \Omega_{q}^2 \Phi_{q}(t) 
				+ \Omega_{q}^2 \int_0^t \!\!d\tau\,m_q(t,\tau)\dot{\Phi}_{q}(\tau) =  0 \,,
			\end{split}
		\end{equation}
		where $\nu_q$ and $\Omega_q$ are characteristic frequencies, and $m_q$ is a memory kernel.
		In simple liquids, the memory kernel is very small, and $\Phi_q(t)$ follows an exponential relaxation, with a typical
		rate related to its diffusion coefficient.
		In complex liquids, the last term of Eq.~(\ref{eqMCT}) grows, and typically leads to the appearance of a plateau in the
		time evolution of $\Phi_q(t)$ that can extend over several decades in time.
		This plateau is related to the so-called "cage effect", namely, the slowing of the average particle's motion due to
		its interaction with its neighbors.

		In the MCT scope, if the plateau of $\Phi_q(t)$ extends to infinite times, the system reaches the solid state.
		Note, however, that such a scenario can never happen in the systems we study here, which are always melted by the external
		shear stress.
		Moreover, as explained above, the granular-liquid regime we study does not extend arbitrarily close to the solid phase.
		We are thus interested only in the emergence of the plateau, due to the steric repulsion between the inelastic hard spheres,
		corresponding to the onset of complex behavior in the liquid.
		Hence, MCT is used precisely in the regime it best describes.
		The full content of the terms in Eq.~(\ref{eqMCT}) is not needed in the rest of this study.
		The interested reader is referred to \cite{Kranz20,Coquand20f} for the details.

		Our main goal is to study the rheology of the system, which we do by studying the evolution of the statistical averages of 
		the components of the stress tensor $\left<\sigma_{\alpha\beta}(t)\right>^{(\dot\gamma)}$.
		Here, the superscript is used to recall that such an average is taken over the configurations of the sheared system, which is
		an out-of-equilibrium system, and is therefore quite challenging to compute.
		A way to simplify its computation is to use the \textit{Integration Through Transients} (ITT) formalism, which relates the
		average in the sheared system $\left<\cdot\right>^{(\dot\gamma)}$ to a statistical average taken in a fictitious reference system,
		noted $\left<\cdot\right>^{(0)}$ \cite{Fuchs02,Fuchs03,Fuchs09}.


		In the context of colloidal suspensions, the reference state is chosen to be the unsheared state, so that 
		the ITT  formalism transforms averages in the out-of-equilibrium sheared system into averages in the unsheared system, which is at equilibrium.
		In the case of granular liquids, we keep this choice of reference state, although because of the dissipative character of the interactions,
		the reference system is still out of equilibrium, and some external source of driving power is needed to maintain a liquid state with nonzero granular temperature.
		More caution should therefore be used in the latter case (see \cite{Kranz20} for details).
		Crucially, both the reference state and the sheared state are stationary states.


		Finally, using the mode coupling approximation, all averages can be projected onto pairs of density operators, so that the ITT equation writes:
		\begin{equation}
		\label{eqITT1}
			\left<\sigma_{\alpha\beta}\right>^{(\dot\gamma)} = \left<\sigma_{\alpha\beta}\right>^{(0)} +\frac{1}{2T}
			\int_0^{+\infty} \!\!\!\!\!dt\int_k\, \mathcal{V}^{\sigma}_{k(-t)}\Phi_{k(-t)}^2(t)\mathcal{W}^\sigma_{k,\alpha\beta}\,,
		\end{equation}
		where $\int_k = \int d^3k/(2\pi)^3$, $T$ is the granular temperature, and the GITT vertices are given by:
		\begin{equation}
		\label{eqVertx}
			\begin{split}
				& \mathcal{V}^\sigma_k = \sum_{\theta,\omega}\kappa_{\theta\omega}\mathcal{V}^\sigma_{k,\theta\omega}
				 = N\left<\kappa:\sigma^{el}\big| \rho_k\rho_{-k}\right>^{(0)} \\
				& \mathcal{W}^\sigma_{k,\alpha\beta} = N\left<\rho_k\rho_{-k}\big| \sigma_{\alpha\beta}\right>^{(0)}/S_k^2 \,,
			\end{split}
		\end{equation}
		where "$:$" denotes a full tensor contraction, $N$ is the number of particles,
		and the superscript "$el$" is here to refer to an equivalent stress tensor in which all collisions are
		considered elastic (the inelastic character of the collisions is already taken into account at the level of the stress tensor appearing in the $\mathcal{W}$ tensor).
		The main difference between the general ITT formalism used to study colloidal suspensions and GITT lies in the vertex $\mathcal{W}$, which
		in particular encodes the dissipative aspect of the collisions.

		Note that due to the mode coupling approximation, most of the time dependence in Eq.~(\ref{eqITT1}) is carried by $\Phi_k(t)$.
		Moreover, since in a liquid (even a complex one) $\Phi_k(t)$ decays at least exponentially fast on a finite time period, the time integral
		is always convergent, which ensures that rheological observables stay finite.
		Notice also that in Eq.~(\ref{eqITT1}), the dynamical structure factor is evaluated at a time dependent wave vector $k(-t)$.
		This wave vector is the wave vector advected by the shear flow, defined as $k(t) = \big(1 + \kappa t)\cdot\mathbf{k}$.
		The presence of the advected wave vector has a physical interpretation and corresponds to the effect of the imposed shear flow on the
		internal dynamics of the system.
		In particular, the advection of particles by the shear flow, by imposing an average velocity profile can facilitate the escape
		of the particles from the cages formed by their neighbors, thereby providing another decay channel for the dynamical structure factor.
		This phenomenon is crucial insofar as it ensures the efficiency of shear melting in granular flows with a high packing fraction
		where the structural relaxations of $\Phi_k(t)$ in the reference system do not trigger any decay.

	\subsection{The GITT vertices}

		The computation of the GITT vertices Eq.~(\ref{eqVertx}) involves \textit{a priori} a high number of different terms.
		However, a clever use of the symmetries can reduce this to a smaller number of defining quantities.

		First, let us introduce the orthogonal projectors respectively onto a given vector $\mathbf{k}$ and orthogonal to it:
		\begin{equation}
		\label{eqProj}
			\begin{split}
				& P_L^{\alpha\beta}(\mathbf{k}) = \frac{k^\alpha k^\beta}{k^2} \\
				& P_T^{\alpha\beta}(\mathbf{k}) = \delta^{\alpha\beta} - \frac{k^\alpha k^\beta}{k^2} \,.
			\end{split}
		\end{equation}
		The microscopic stress tensor can be decomposed on this projector's basis as:
		\begin{equation}
		\label{eqsLT0}
			\sigma_{\alpha\beta}(\mathbf{k}) = \sigma_L(k)\, P_L^{\alpha\beta}(\mathbf{k}) + \sigma_T(k)\, P_T^{\alpha\beta}(\mathbf{k})\,,
		\end{equation}
		so that the computation of the vertices reduces to the evaluation of $\left<\rho_k\rho_{-k}\big|\sigma_L\right>$, $\left<\rho_k\rho_{-k}\big|\sigma_T\right>$,
		and their analog in reverse order (in granular liquids, the time-reversal symmetry is broken due to the dissipative collisions).
		This yields, for example \cite{Kranz20},
		\begin{equation}
			\label{eqsLT}
			\begin{split}
				&N\left<\rho_k\rho_{-k}\big| \sigma_{L}\right>^{(0)} = \frac{1+\varepsilon}{2}\, T\big[-kS_k' + S_k -S_k^2\big] \\
				&N\left<\rho_k\rho_{-k}\big| \sigma_{T}\right>^{(0)} = \frac{1+\varepsilon}{2}\, T\big[S_k -S_k^2\big] \,,
			\end{split}
		\end{equation}
		where $S_k' = dS_k/dk$.
		Note in particular that the contribution of the dissipative collisions factors out, and has a well-defined elastic limit $\varepsilon\rightarrow1$,
		yielding back the well-known results in the MCT study of colloids \cite{Fuchs02}.
		A detailed computation of these averages can be found in \cite{Kranz13}.

		Before going further, let us introduce some more notations.
		Indeed, it can be seen in Eq.~(\ref{eqITT1}) that in a general flow, many terms such as those in Eq.~(\ref{eqsLT}) would be present,
		so that it is useful to define generic scalar quantities which do not depend on $\varepsilon$, for example, and can therefore be used in the definition
		of both $\mathcal{V}$ and $\mathcal{W}$.
		Also, in order to perform the 3D momentum integral in Eq.~(\ref{eqITT1}), it is useful to decompose the vertices according to their components along
		$k$ rather than along the orthogonal projectors of Eq.~(\ref{eqProj}).
		We therefore define
		\begin{equation}
		\label{eqsDO}
			\begin{split}
				& \sigma_\perp = T\big[S_k - S_k^2\big] \\
				& \Delta\sigma = - T S_k' \,,
			\end{split}
		\end{equation}
		which finally allow us to rewrite the vertices as
		\begin{equation}
		\label{eqVW}
			\begin{split}
				& \mathcal{V}_{k,\alpha\beta}^\sigma = \hat{k}_\alpha\hat{k}_\beta\,k\Delta\sigma + \delta_{\alpha\beta} \sigma_\perp \\
				& \mathcal{W}_{k,\alpha\beta}^\sigma = \frac{1+\varepsilon}{2S_k^2}\big[
					\hat{k}_\alpha\hat{k}_\beta\,k\Delta\sigma + \delta_{\alpha\beta} \sigma_\perp\big] \,,
			\end{split}
		\end{equation}
		where $\hat{k}_\alpha=k_\alpha/k$.
		The additional $k$ factor with $\Delta\sigma$ comes from its definition in Eq.~(\ref{eqsDO}) relative to Eq.~(\ref{eqsLT}).

		Finally, we have decomposed the vertices in such a way that all their anisotropic part appears explicitly
		It is thus possible to perform the integral over the angular variables in Eq.~(\ref{eqITT1}), which leads to:
		\begin{equation}
		\label{eqITT2}
			\begin{split}
				\left<\sigma_{\alpha\beta}\right>^{(\dot\gamma)} = &\left<\sigma_{\alpha\beta}\right>^{(0)} \\
				&+\frac{1}{T}\int_0^{+\infty} \!\!\!\!\!dt\int_0^{+\infty} \!\!\!\!\!dk\,k^2 \sum_{\omega,\theta}\kappa_{\omega\theta}
				\mathbb{J}_{\alpha\beta}^{\quad\omega\theta}\,\Phi_{k(-t)}^2(t)\,,
			\end{split}
		\end{equation}
		where $\mathbb{J}_{\alpha\beta}^{\quad\omega\theta}$ is the result of the integral over the angular variables of the product of both GITT vertices.
		Its full expression is given in the Appendix \ref{AJ} for symmetric $\kappa$s.
		Our expressions can be easily generalized to the non symmetric case with the same reasoning.

		The GITT formula (\ref{eqITT2}) can also be written as
		\begin{equation}
		\label{eqL}
			\left<\sigma_{\alpha\beta}\right>^{(\dot\gamma)} = \left<\sigma_{\alpha\beta}\right>^{(0)} +\sum_{\theta\omega}\kappa_{\theta\omega}
			\Lambda_{\alpha\beta\theta\omega}(\dot\gamma)\,,
		\end{equation}
		where the viscosity tensor $\Lambda_{\alpha\beta\theta\omega}$ plays a role analogous to that of the elasticity tensor in the theory of elasticity.
		Let us stress however that this is not to be confused with a linear approximation of the relation between $\left<\sigma_{\alpha\beta}\right>^{(\dot\gamma)}$
		and $\kappa$; this has been emphasized in Eq.~(\ref{eqL}) by the explicit mention of the fact that $\Lambda_{\alpha\beta\theta\omega}$ is a function
		of $\dot\gamma$.
		The explicit expression of $\Lambda_{\alpha\beta\theta\omega}$ as well as its decomposition in powers of $\kappa$ is discussed in the next section.

		All in all, we presented the extension of the GITT equations to the case of a general incompressible stationary flow.
		We wrote it in a form of a double (convergent) integral depending on the flow tensor $\kappa$ and a vertex tensor $\mathbb{J}$,
		which explicit expression in terms of the only two scalars $\Delta\sigma$ and $\sigma_\perp$ --- defined in Eq.~(\ref{eqsDO}) --- is given
		in the equations (\ref{AeqJ1}), (\ref{AeqJ2}), (\ref{AeqJ3}), and (\ref{AeqJ4}).
		The general GITT equation (\ref{eqITT2}) can then be integrated numerically for any given granular-liquid flow, thereby giving access to the
		macroscopic averages of all the components of the stress tensor in the stationary state in the sheared system.

\subsection{The viscosity tensor $\Lambda_{\alpha\beta\theta\omega}$}
\label{AB}

	Let us discuss in more detail the viscosity tensor $\Lambda$ defined in Eq.~(\ref{eqL}) which is the analog of the elasticity tensor of solids and describes how the stress
	tensor is defined from the flow tensor in presence of shear.
	Indeed, within the GITT formalism, this tensor can be expressed explicitly in terms of a small number of elementary integrals,
	which allow us to get access to its finer structure.
	The following reasoning is fully general and does not suppose that the flow tensor $\kappa$ is symmetric.

	First, notice that the tensorial structure of $\Lambda$ is directly caused by that of $\mathbb{J}$, more precisely, it is inherited from the momentum structure of the
	GITT vertices $\mathcal{V}$ and $\mathcal{W}$, itself coming from the decomposition of the $\sigma$ operator into its components longitudinal and transverse to $k$
	; see Eq.~(\ref{eqsLT0}).
	It is therefore the definition of the microscopic $\sigma$ itself that guarantees that $\Lambda$ is a tensor of rank four.

	The tensorial structure of $\mathcal{V}$ and $\mathcal{W}$ has been expressed in Eq.~(\ref{eqVW}) in terms of $\sigma_\perp$ and $\Delta\sigma$.
	In terms of these variables, we can reduce the tensorial structure of the integrand to

	\begin{equation}
	\label{eqAB1}
		\begin{split}
			\mathcal{V}^\sigma_{k,\theta\omega}\mathcal{W}^\sigma_{k,\alpha\beta}\propto & \delta_{\alpha\beta}\delta_{\theta\omega}\sigma_{\perp}^2
			+ \delta_{\theta\omega}\hat{k}_\alpha\hat{k}_\beta k \sigma_\perp\Delta\sigma \\
			& + \delta_{\alpha\beta}\hat{k}_\theta(-t)\hat{k}_\omega(-t)k(-t) \sigma_\perp\Delta\sigma\\
			&+ \hat{k}_\alpha\hat{k}_\beta\hat{k}_\theta(-t)\hat{k}_\omega(-t) k k(-t)\Delta\sigma^2\,.
		\end{split}
	\end{equation}

	Let us examine these terms one by one.

	The two first terms of Eq.~(\ref{eqAB1}) are proportional to $\delta_{\theta\omega}$.
	Hence they never contribute for incompressible flows.
	Using the formula

	\begin{equation}
	\label{eqAB2}
		\int_k \,k_ik_j\,f(k^2) = \frac{\delta_{ij}}{3}\int_k f(k^2)\,,
	\end{equation}
	true for any smooth function $f$, we can reexpress the two first terms of Eq.~(\ref{eqAB1}) as $\delta_{\alpha\beta}\delta_{\theta\omega}\big(\mathcal{B}_0^{comp}
	+ \mathcal{B}_1^{comp}\big)$, where we defined the two following scalars:
	\begin{equation}
	\label{eqAB3}
		\begin{split}
			& \mathcal{B}_0^{comp} = \int_0^{+\infty}dt \int_k \frac{1+\varepsilon}{2S_k^2} \Phi_{k(-t)}^2(t)\frac{\sigma_\perp^2}{2T} \\
			& \mathcal{B}_1^{comp} = \int_0^{+\infty}dt \int_k \frac{1+\varepsilon}{2S_k^2} \Phi_{k(-t)}^2(t)\frac{k\sigma_\perp\Delta\sigma}{6T} \,.
		\end{split}
	\end{equation}

	For the next terms, we need to remember that $k_i(t) = (\delta_{ij} + \kappa_{ij}t)k_j$, where the sum over repeated indices is implicit.
	Then, the third term of Eq.~(\ref{eqAB1}) can be written $\delta_{\theta\omega}\delta_{\alpha\beta}\mathcal{B}_2^{comp} - \delta_{\alpha\beta}D_{\theta\omega}\mathcal{B}_P^1
	+ \delta_{\alpha\beta}\kappa_{\theta i}\kappa_{i\omega}\mathcal{B}_P^2$, where we defined the following quantities:
	\begin{equation}
	\label{eqAB4}
		\begin{split}
			& \mathcal{B}_2^{comp} = \int_0^{+\infty}dt \int_k \frac{1+\varepsilon}{2S_k^2} \Phi_{k(-t)}^2(t)
			\frac{k^2\sigma_\perp\Delta\sigma}{6k(-t)T} \\
			& \mathcal{B}_1^{P} =    \int_0^{+\infty}dt \int_k \frac{1+\varepsilon}{2S_k^2} \Phi_{k(-t)}^2(t)
			\frac{k^2\sigma_\perp\Delta\sigma}{6k(-t)T}\,t \\
			& \mathcal{B}_2^{P} =    \int_0^{+\infty}dt \int_k \frac{1+\varepsilon}{2S_k^2} \Phi_{k(-t)}^2(t)
			\frac{k^2\sigma_\perp\Delta\sigma}{6k(-t)T}\,t^2 \,.
		\end{split}
	\end{equation}
	Again, the term proportional to $\mathcal{B}_2^{comp}$ does not yield any contribution to $\left<\sigma_{\alpha\beta}\right>^{(\dot\gamma)}$ for compressible flows.
	Moreover, since the other terms are proportional to $\delta_{\alpha\beta}$, they do not contribute to the shear stress $\sigma_0$ for the same reason, hence the $P$
	superscript.

	For the last term, we need the formula
	\begin{equation}
	\label{eqAB5}
		\int_k \,k_ik_jk_ak_b\,f(k^2) = \frac{X_{ijab}}{15}\int_k f(k^2)\,,
	\end{equation}
	where $X$ is the fully symmetric tensor of rank four expressed in terms of $\delta$-symbols:
	\begin{equation}
	\label{eqABX}
		X_{ijab} = \delta_{ij}\delta_{ab} + \delta_{ia}\delta_{jb}+ \delta_{ib}\delta_{ja}\,.
	\end{equation}
	We also define the $Y^{n}_{ijab}$ tensors which are the symmetric combinations of $n$ $\kappa$ tensors and $\delta$-symbols:
	\begin{equation}
	\label{eqABY}
		\begin{split}
			& Y_{ijab}^{1} = \delta_{ij}D_{ab} + \kappa_{ai}\delta_{bj} + \kappa_{bj}\delta_{ai} + \kappa_{aj}\delta_{bi} + \kappa_{bi}\delta_{aj} \\
			& Y_{ijab}^{2} = \delta_{ij}\kappa_{ac}\kappa_{cb} + \kappa_{ai}\kappa_{bj} + \kappa_{bi}\kappa_{aj} \,,
		\end{split}
	\end{equation}
	and the following scalars:
	\begin{equation}
	\label{eqAB5}
		\begin{split}
			& \mathcal{B}_X^{\sigma} = \int_0^{+\infty}dt \int_k \frac{1+\varepsilon}{2S_k^2} \Phi_{k(-t)}^2(t)
			\frac{k^3\Delta\sigma^2}{30k(-t)T} \\
			& \mathcal{B}_1^{\sigma} = \int_0^{+\infty}dt \int_k \frac{1+\varepsilon}{2S_k^2} \Phi_{k(-t)}^2(t)
			\frac{k^3\Delta\sigma^2}{30k(-t)T}\,t \\
			& \mathcal{B}_2^{\sigma} = \int_0^{+\infty}dt \int_k \frac{1+\varepsilon}{2S_k^2} \Phi_{k(-t)}^2(t)
			\frac{k^3\Delta\sigma^2}{30k(-t)T}\,t^2 \,,
		\end{split}
	\end{equation}
	in terms of which $\Lambda$ finally results in:
	\begin{equation}
	\label{eqABL}
		\begin{split}
			\Lambda_{\alpha\beta\theta\omega} &= \big(\mathcal{B}_0^{comp} + \mathcal{B}_1^{comp}+ \mathcal{B}_2^{comp}\big)\delta_{\alpha\beta}\delta_{\theta\omega} \\
			&- \mathcal{B}_1^P\delta_{\alpha\beta}D_{\theta\omega} + \mathcal{B}_2^P\delta_{\alpha\beta}\kappa_{\theta i}\kappa_{i \omega}  \\
			&+ \mathcal{B}_X^\sigma X_{\alpha\beta\theta\omega} + \mathcal{B}_1^\sigma Y^1_{\alpha\beta\theta\omega} + \mathcal{B}_2^\sigma Y^2_{\alpha\beta\theta\omega}\,.
		\end{split}
	\end{equation}
	This formula can be read as follows: the first line are terms that contribute only for compressible flows, in the second line are the terms that contribute
	only to the pressure, and the last line are terms that contribute to the shear stress.
	Terms are also classified according to the number of $\kappa$ terms involved.
	In particular, notice that four of those terms involve at least one power of $\kappa$, thereby showing that the GITT approximation is not a low $\dot\gamma$ expansion.
	Additional non polynomial $\dot\gamma$ dependence is also present in the dynamical structure factor $\Phi_{k(-t)}(t)$ that accounts for the effect of advection.

	It can also be inferred from Eq.~(\ref{eqABL}) that $\Lambda$ involves at best two powers of the $\kappa$ tensor, which is consistent with the expressions of the toy model
	where the only $\mathcal{K}_n$ integrals present in the expression of the rheological observables have $n\leqslant2$.
	This can be traced back to the fact that the $\mathcal{V}$ vertex, which is the time-dependent vertex, is expressed as a statistical average involving the stress
	tensor, and can therefore involve at most two $k_i(-t)$ components, and to our approximation of the time dependence of the wave vector.

\section{Rheology in the GITT framework}

	\subsection{Rheological observables}

		In order to compare different flow configurations, we first need to build from the matrix $\kappa$ a scalar quantity, the
		\textit{effective strain rate} $\dot{\gamma}^{eff}$, defined by:
		\begin{equation}
		\label{eqgeff}
			\dot{\gamma}^{eff} = \sqrt{\frac{1}{2}\,D:D} \,,
		\end{equation}
		where $D$ is the symmetrized flow matrix $D = \kappa + \kappa^T = \nabla\cdot\mathbf{v} + \nabla\cdot\mathbf{v}^T$.
		In particular, in the case of simple shear flow, $\dot{\gamma}^{eff} = \dot\gamma$, but this does not hold for a general flow.

		Then, we also need to extract from the nine components of $\sigma$ fundamental scalar quantities that can be used to compare
		different flow configurations.
		In order to do so, we decompose as usual the stress tensor in a diagonal and a deviatoric component:
		\begin{equation}
		\label{eqsigma}
			\sigma_{\alpha\beta} = P\,\delta_{\alpha\beta} + \eta \, D_{\alpha\beta}\,,
		\end{equation}
		which defines the pressure $P$, as well as the shear viscosity $\eta$ of the liquid. 
		Note that this decomposition has a genuinely fundamental character since it corresponds to the identification of the
		spin 0 and spin 2 components of $\sigma$ on the irreducible representation of the SO$(3)$ group of symmetry.
		It is analogous to the decomposition of the stress tensor of a solid along a bulk and a shear modulus.

		The decomposition (\ref{eqsigma}) is also useful to extend to general incompressible flows the well-known relationship
		between $\eta$, the shear rate and the shear stress.
		Whereas in the case of a simple shear flow, the shear stress corresponds to the only nonzero component of $\sigma_{\alpha\beta}$
		out of the diagonal, the definition of such a quantity in a more general flow configuration may not be so obvious.

		First, remark that $\eta$ can be isolated by a full contraction of $\sigma$ with the tensor $D$ because the flow is incompressible: $\sigma:D = \eta D:D$.
		Then, since $\sigma$ is symmetric, $\sigma:D = 2\sigma:\kappa$, so that finally $\sigma:\kappa = \eta \big(\dot{\gamma}^{eff}\big)^2$.
		Thus, we can define a (scalar) shear stress $\sigma_0$ by
		\begin{equation}
		\label{eqs0}
			\sigma_0 \,=\, \frac{\sigma:\kappa}{\dot{\gamma}^{eff}} \,=\, \eta\,\dot{\gamma}^{eff} \,.
		\end{equation}
		Note that this definition is still valid in the case of simple shear flows.

		Finally, all these definitions can be used to extend the definition of the effective friction coefficient $\mu$.
		This quantity is originally defined as a ratio between a normal and a tangential force applied to an elementary cell of
		the fluid.
		However, although such a definition is well suited to the simple shear flow, it is a bit too close to the original law of Coulomb
		to be directly used in more general flow configurations.
		In this study, we extended the definition of the effective friction coefficient from the decomposition (\ref{eqsigma}) as
		\begin{equation}
		\label{eqmu}
			\mu = \frac{\sigma_0}{P} \,,
		\end{equation}
		which as explained above yields the usual result in the case of simple shear flow.

		All in all, in the following analysis a given flow configuration will be characterized by three scalar quantities,
		$P$, $\sigma_0$, and $\mu$, which capture the fundamental properties of the flow rheology, and can be compared in different
		flow configurations.

	\subsection{Reduction of the ITT integrals}

		The computation of the rheological observables from Eq.~(\ref{eqITT2}) is a rather involved task.
		In order to get a better understanding of the physical processes at play in the system, it is useful to reduce it to simpler toy models,
		as was done in \cite{Coquand20g}.

		Indeed, as far as the rheology is concerned, the central quantities are global quantities defined at the macroscopic scale of the system.
		We can therefore get rid of most of the momentum dependence, which describes finer structures.
		More precisely, let us write $\Phi_k(t)$ in the Vineyard approximation \cite{Vineyard58} and reduce the self-interacting part of the dynamical
		structure factor to a Gaussian function of the mean-squared displacement (MSD) $\Delta r^2$:
		\begin{equation}
		\label{eqPhi}
			\Phi_k(t) \simeq S_k \, e^{-k^2\Delta r(t)^2}\,.
		\end{equation}
		The static structure factor $S_k$ is merely an oscillating function of order one, which role can be overlooked at lowest order.
		The controlling factor in Eq.~(\ref{eqPhi}) is the second one.
		In particular, in a liquid, the MSD follows the law of diffusion, so that $\Phi_k(t\rightarrow+\infty) \sim \exp(-\Gamma_k t)$, where $\Gamma_k$
		is some ($k$-dependent) decay rate.
		Above the ideal MCT glass transition however, the MSD saturates to some constant value due to the cage effect, so that $\Phi_k$ also saturates and
		never decays to 0.
		We can therefore propose the rather bold ansatz $\Phi_k(t)\simeq \exp(-\Gamma t)$ to replace the solution to the MCT equation (\ref{eqMCT})
		in the ITT integral of Eq.~(\ref{eqITT2}), which as we have shown captures the main phenomenology of MCT.

		However, we must not forget that in Eq.~(\ref{eqITT2}) $\Phi_k(t)$ is evaluated in the \textit{advected} wave vector $k(-t)$.
		This has a major importance since in Eq.~(\ref{eqPhi}) replacing $k^2$ by the ever increasing $k(-t)^2$ provides a new decay channel for the
		dynamical structure factor, which ensures that even above the ideal MCT glass transition the system remains in a liquid state (it is shear molten).
		For a general shear flow, the time-dependent wave vector is
		\begin{equation}
		\label{eqq}
			\begin{split}
				k^2(-t) &= \mathbf{k}\cdot\big(1 - D\,t + \kappa^T\cdot\kappa\,t^2\big)\cdot\mathbf{k} \\
				&= k^2 -2t\Big[\sum_\alpha \text{Tr}(\kappa)k_\alpha^2 + \sum_{\alpha\neq\beta}\kappa_{\alpha\beta}k_\alpha k_\beta\Big] \\
				& \ +t^2\Big[\sum_{\alpha,\beta,\theta}\kappa_{\alpha\theta}\kappa_{\beta\theta}k_\alpha k_\beta\Big] \,.
			\end{split}
		\end{equation}
		Now examine the second term of Eq.~(\ref{eqq}).
		There are two terms in the bracket: the first one vanishes because the flow is incompressible (Tr$(\kappa)=0$), and the second one does not contribute
		to $k^2$ on average.
		Finally, $\left<k^2(-t)\right> - k^2\propto t^2$.
		Therefore, the effect of advection on the wave vector in the ITT integral can be taken into account by a Gaussian factor $\exp(-\dot\gamma^2 t^2/\gamma_c^2)$
		where $\gamma_c$ is a typical strain scale.
		This Gaussian factor is comparable to the screening factor used to study the rheology of colloidal suspensions with schematic MCT models
		\cite{Fuchs02,Fuchs03,Fuchs09,Brader09}, although in that case a Lorentzian profile was chosen.
		The precise form of the profile is mostly irrelevant at our level of approximation \cite{Coquand20g}.

		Our ansatz, modified by the Gaussian advection profile, can then be inserted into Eq.~(\ref{eqITT2}).
		Interestingly, since the wave vector dependence has been left aside, the whole $k$ integral, which contains much of the complexity of the problem through
		the combination of structure factors $S_k$ reduces to a mere constant prefactor.
		The remaining $t$ integrals can then be expressed in terms of a linear combination of integrals from the following family:
		\begin{equation}
		\label{eqK}
			\mathcal{K}_i = \int_0^{+\infty}\dot\gamma dt\,\big(\dot\gamma t\big)^i\,e^{-2\Gamma t-2\dot\gamma^2 t^2/\gamma_c^2}\,.
		\end{equation}
		As we have shown in \cite{Coquand20g}, the integral $\mathcal{K}_0$ can be computed exactly, and can be estimated to a very good precision by:
		\begin{equation}
		\label{eqK0}
			\mathcal{K}_0= \frac{\overline{\gamma}_c}{2(1 + \overline{\gamma}_c\,u)}\,,
		\end{equation}
		where $\overline{\gamma}_c = \gamma_c\,\sqrt{\pi/2}$ and $u = \Gamma/\dot\gamma$ is the ratio of the two rates that compete
		for the control of the final decay of $\Phi_{k(-t)}(-t)$.
		The further integrals are related to $\mathcal{K}_0$ by
		\begin{equation}
		\label{eqKi}
			\mathcal{K}_i = \left(-\frac{\dot\gamma}{2}\frac{d}{d\Gamma} \right)^i\mathcal{K}_0 \,.
		\end{equation}
		Finally, we have shown that it is possible to reduce the ITT integral (\ref{eqITT2}) to a linear combination of $\mathcal{K}_i$ integrals,
		defined by the vertex tensor $\mathbb{J}_{\alpha\beta}^{\quad\theta\omega}$.
		These integrals can be reduced to rational fractions of one variable, $u$, which is the ratio of the possible rates controlling the final decay of $\Phi_k(t)$.
		This constitutes the so-called two time scales toy model which captures successfully the lowest order behavior of rheological observables
		as has been shown in \cite{Coquand20g}.
		Concrete applications will be given in the following of the paper.

		More subtle variations, such as the $\mu(\mathcal{I})$ law (\ref{eqMuI}), require a more precise model to be properly described.
		A way to do so is to enrich the ansatz we used to describe $\Phi_k(t)$.
		Indeed, we have so far reduced the properties of $\Phi$ to the nature of the time scale responsible for its decay, which can be either the structural relaxations,
		with a rate $\Gamma$, or the shear advection, with a rate $\dot\gamma$.
		But another salient feature of $\Phi_k(t)$ is the existence or absence of a plateau corresponding to the influence of cage effect in the complex liquid phase.
		This is related to the fact that a precise description of granular-liquid flows involves not one but two dimensionless numbers: the P\'eclet Pe number, related to the
		ratio of the diffusion and advection time scales, and the Weissenberg number Wi, which is a ratio of structural relaxation and advection time scales.
		The P\'eclet number can be expressed from the collision frequency $\omega_c$ as Pe$=\dot\gamma/\omega_c$; it is the ratio of a time scale associated with the
		microscopic motion of the particles with the advection time scale $\dot\gamma^{-1}$.
		It is therefore proportional to the inertial number $\mathcal{I}$.
		On the other hand, Wi =$1/u=\dot\gamma/\Gamma$.
		The plateau of the cage effect typically develops between the microscopic time scale and the final decay of $\Phi(t)$.

		In a Newtonian liquid, the decay of $\Phi$ is given by $\Gamma\propto\omega_c$, so that Wi $\propto$ Pe, the microscopic time scale corresponds to the decay
		of $\Phi$, and there is no plateau.
		In a complex liquid, the structural relaxation and microscopic time scales decouple, and the plateau develops in between.

		A simple way to incorporate this into the toy model is to use a two-step decay ansatz for $\Phi_k(t)$ \cite{Levesque73,Coquand20g}:
		\begin{equation}
		\label{eqPhi2s}
			\Phi_k(t)=\lambda^{(1)}\exp(-\Gamma^{(1)}t) +\lambda^{(2)}\exp(-\Gamma^{(2)}t)  \,,
		\end{equation}
		where of course the advection factor has to be added when replacing $\Phi_{k(-t)}(t)$ in Eq.~(\ref{eqITT2}).
		From the two decay rates of Eq.~(\ref{eqPhi2s}) we can define $u^{(1)} = \Gamma^{(1)}/\dot\gamma\propto1/$Pe and $u^{(2)} = \Gamma^{(2)}/\dot\gamma\propto1/$Wi.
		This constitutes the three time scales toy model.
		By linearity of the ITT integrals, the general form of the components of the stress tensor in the three time scales toy model is the same than in the
		two time scales toy model, except that each term appears twice, once with each variable $u^{(i)}$, with appropriate set of constants.
		Strictly speaking the ITT integrand is not linear in $\Phi$. However, 	since the toy models mostly make sense for well separated time scales, the
		square of the exponential sum has still the form of a two step decay, so that $\Phi^2$ still has the form Eq.~(\ref{eqPhi2s}).

	\subsection{Application to the simple shear case}

		Let us now recall how this formalism applies in the case of simple shear flows.
		This will also provide a useful point of comparison for our further investigations.
		The interested reader can find the details of the derivation in \cite{Coquand20f,Coquand20g}.

		The simple shear flow is defined from the flow matrix $\kappa_{ij}^{ss} = \dot\gamma \delta_{ix}\delta_{jy}$.
		In that case, Eq.~(\ref{eqgeff}) yields $\dot\gamma^{eff} = \dot\gamma$, Eq.~(\ref{eqs0}) yields $\sigma_0^{ss}=\left<\sigma_{xy}\right>$,
		and as usual $P^{ss}=$Tr$(\sigma)/3$ and $\mu^{ss} = \sigma^{ss}_0/P^{ss}$.

		The GITT vertex tensor cannot be directly gotten from the results in the Appendix \ref{AJ} since $\kappa$ is not symmetric.
		For the shear stress, it is given by
		\begin{equation}
		\label{eqJs0}
			\mathbb{J}_{xy}^{\quad xy} = \frac{1}{60\pi^2} \frac{1+\varepsilon}{2S_k^2}\frac{k^3}{k(-t)}\Delta\sigma^2\,.
		\end{equation}
		Hence, $\sigma_0^{ss}$ is proportional to $\mathcal{K}_0$.
		By convention, we call $S_1$ the prefactor accounting for the whole $k$ integral.
		Thus,
		\begin{equation}
		\label{eqs0ss0}
			\sigma_0^{ss} = S_1\,\mathcal{K}_0 \,.
		\end{equation}
		The pressure vertex tensor is given by
		\begin{equation}
		\label{eqJP}
			\begin{split}
				\mathbb{J}_{xy}^{\quad xx} + \mathbb{J}_{xy}^{\quad yy} +\mathbb{J}_{xy}^{\quad zz} =&
				\frac{1+\varepsilon}{2S_k^2}\frac{k^3}{k(-t)}\bigg\{\frac{\sigma_\perp\Delta\sigma}{4\pi^2}\big(\dot\gamma t\big) \\
				&+\frac{k\,\Delta\sigma^2}{12\pi^2}\big(\dot\gamma t\big)\bigg\}\,.
			\end{split}
		\end{equation}
		The pressure is therefore expressed in terms of the integral $\mathcal{K}_1$.
		In the second term of Eq.~(\ref{eqJP}), the wave vector integral includes a $\Delta\sigma^2$ term, and it is therefore proportional to $S_1$.
		The first term contains an integral of type $\sigma_\perp\Delta\sigma$, which we use to define a new constant $S_0$.
		Finally,
		\begin{equation}
		\label{eqPss0}
			P^{ss} = P_0 + \Delta P^{ss} = P_0+ \left(S_0 + \frac{5}{3}S_1\right)\mathcal{K}_1 \,,
		\end{equation}
		where $P_0$ is the pressure in the reference state, which is not sheared.
		It corresponds to the pressure of a hard sphere fluid at the same packing fraction and granular temperature (see \cite{Coquand20f} for a detailed discussion).

		Now let us examine the two time scales toy model.
		In that model, the rheological observables are given by a competition between the time scales of structural relaxation $t_\Gamma = 1/\Gamma$ and
		advection $t_\gamma=1/\dot\gamma$ that compete for the control of the decay of $\Phi(t)$.
		They are thus functions of $u$ given by the simplified version of the $\mathcal{K}_i$ integrals derived from Eq.~(\ref{eqK0}).
		The shear stress is
		\begin{equation}
		\label{eqs0ss1}
			\sigma_0^{ss} = \frac{\sigma_y^{ss}}{1 + \overline{\gamma}_c u} \,,
		\end{equation}
		where $\sigma_y^{ss} = S_1 \overline{\gamma}_c$.
		This result is interpreted as follows: in the Newtonian regime, the decay of $\Phi$ is controlled by structural relaxations; namely, the decay is caused by the
		rate $\Gamma$ independent on the value of $\dot\gamma$, hence $t_\Gamma\ll t_\gamma$.
		This is the structural relaxation-dominated regime.
		In that regime $u\gg1$, so that $\sigma_0^{ss}\simeq \eta_0^{ss}\,\dot\gamma$, which is the constitutive equation of a Newtonian liquid of shear viscosity
		$\eta_0^{ss} = \sigma_y^{ss}/(\Gamma\overline{\gamma}_c)$.

		On the other hand, if $t_\gamma\ll t_\Gamma$, the decay is caused by advection, this is the advection dominated regime.
		In that regime, $\sigma_0^{ss}\simeq \sigma_y$, which is the constitutive equation of a yielding fluid.
		In particular, the analog of Hooke's law in the yielding regime leads to the identification of the shear stress $G_\infty$ of the liquid
		from $\sigma_y^{ss} = G_\infty^{ss} \,\overline{\gamma_c}$, so that the shear modulus of the yielding fluid is $G_\infty = S_1/2$, which gives a physical meaning
		to this constant.

		If we further impose that all the energy brought to the system by the shear is dissipated by the collisions, the system gets into the Bagnold regime.
		In that case we need to make explicit the dependence of $\sigma_y$ on the granular temperature, which finally yields $\sigma_0^{ss}\simeq B \dot\gamma^2$,
		where $B=\hat\sigma_y^3/\Gamma_d$ is the Bagnold coefficient, $\hat\sigma_y=\sigma_y/T$, and $\Gamma_d$ is the dimensionless rate of energy dissipated
		by the inelastic collisions (see \cite{Coquand20g} for more details).
		Importantly, both the Bagnold and the yielding regime are advection-dominated regimes, which means that $u\ll1$.

		A similar study of the pressure yields
		\begin{equation}
		\label{eqPss1}
			P^{ss} = P_0 + \frac{P_1^{ss}}{\big(1 + \overline{\gamma}_c u\big)^2} \,,
		\end{equation}
		where $P_1^{ss} = (S_0+5/3 S_1)\overline{\gamma}_c^2/4$ is the overpressure caused by the dynamics of the sheared liquid in the advection dominated
		regime and $P_0$ is the pressure of the unsheared fluid.
		In the structural relaxation dominated regime, the second term in Eq.~(\ref{eqPss1}) is proportional to $\dot\gamma^2$, so that $\Delta P \ll P_0$.
		This is consistent with the fact that the overpressure effect is expected to be negligible in Newtonian liquids.
		In the advection dominated regime --- yielding and Bagnold regimes --- however, the correction to the pressure is significant.
		Numerical estimations from the GITT formula (\ref{eqITT2}) indicate that in that case $\Delta P$ and $P_0$ have the same order of magnitude
		(see \cite{Coquand20f} for a detailed study of the pressure term).

		Finally, $\mu$ is the ratio of Eq.~(\ref{eqs0ss1}) and Eq.~(\ref{eqPss1}).
		It therefore takes the form of a Pad\'e approximant with a numerator being a polynomial of degree 1, and a denominator
		being a polynomial of degree 2, which we write as $P[1/2]$.
		However, as can be guessed from Eq.~(\ref{eqMuI}), not all the constants defined there are important.
		In practice, $\Delta P$ can be approximated further by $\Delta P\simeq P_1'/(1 + \overline{\gamma}_c u)$, which defines the constant $P_1'$.
		In that case, $\mu$ takes the form
		\begin{equation}
		\label{eqMuss0}
			\mu^{ss} = \frac{M_1}{1 + M_2\, u}\,,
		\end{equation}
		with $M_1 = \sigma_y/(P_0 + P_1')$ and $M_2 = \overline{\gamma}_c P_0/(P_0 + P_1')$.
		Hence, in the structural relaxation dominated regime, $\mu^{ss}\propto\dot\gamma$, so that $\mu\ll1$ as expected in the Newtonian regime.
		In the advection dominated regime, $\mu^{ss}\simeq M_1$, it saturates to a finite value.
		This behavior can be observed when solving the GITT equation (\ref{eqITT2}) numerically \cite{Coquand20g}.

\begin{figure}
				\begin{center}
					\includegraphics[scale=0.5]{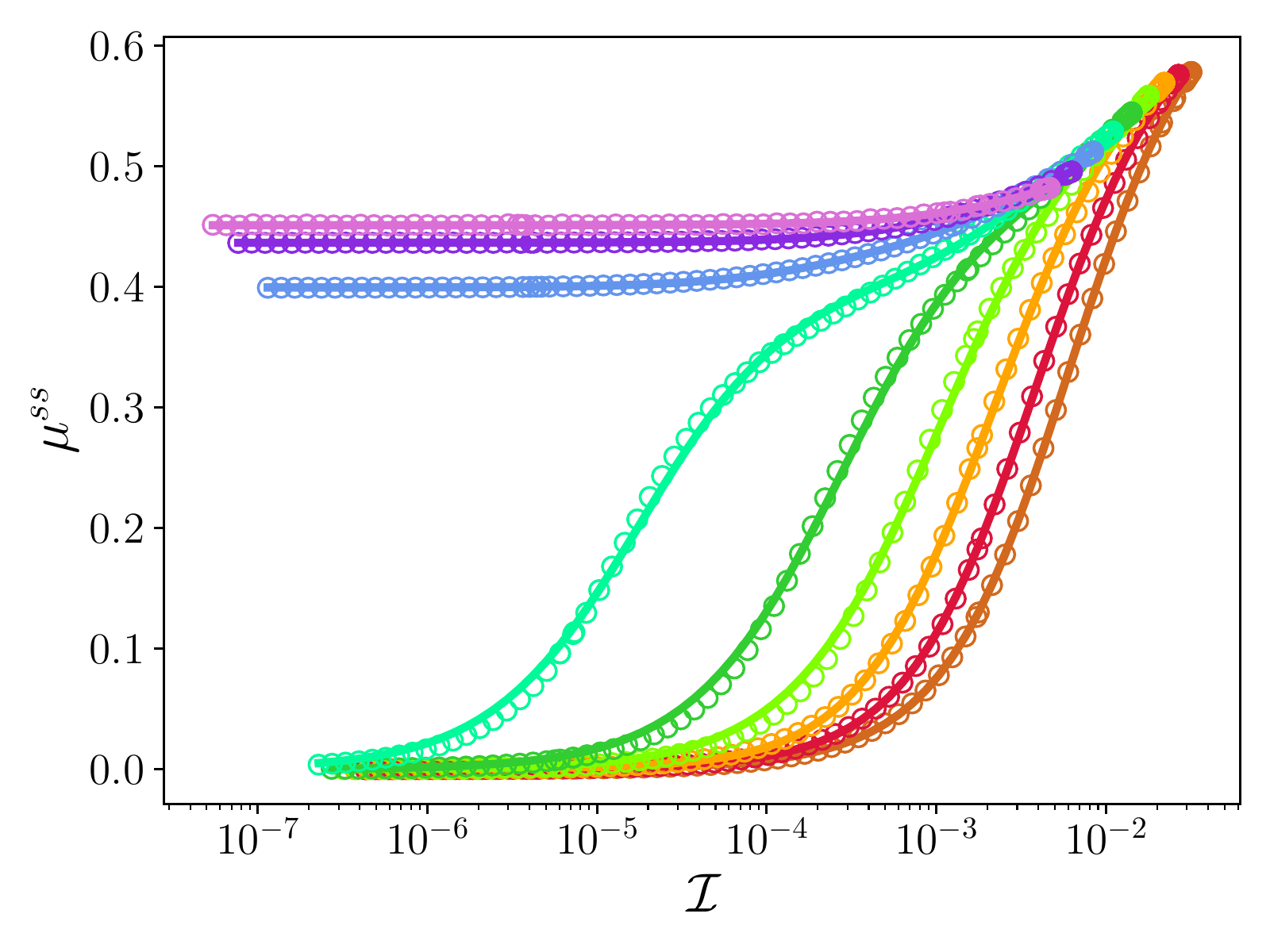}
				\end{center}
				\caption{Evolution of the effective friction coefficient in the simple shear flow with the inertial number for various packing fractions from
				$\varphi=0.42$ (bottom) to $\varphi = 0.58$ (top).
				The open circles are the numerical data gotten from the full GITT equation (\ref{eqITT2}); the full lines are curve fitting with help
				of the toy model expressions (\ref{eqMuss1}).}
				\label{FigMuss}
			\end{figure}

		In order to get more precision, it is instructive to apply the three-time scales toy model.
		Here, in addition to the time scale controlling the final decay, we also want to examine the time scales delimiting the possible cage effect plateau
		in the evolution of $\Phi(t)$.
		This leads us to introduce the time scale associated with the microscopic motion of the granular particles, which is simply the time scale associated with the
		motion of particles in a pressure field $t_m = d/\sqrt{P/\rho}$, $d$ being the diameter of the particles.
		In particular, $\mathcal{I} = t_m/t_\gamma$.
		From the previous analysis we deduce that rheological observables are then functions of two time scale ratios: $u^{(1)} = t_\gamma/t_m\propto 1/\mathcal{I}$,
		and $u^{(2)} = t_\gamma/t_\Gamma\propto 1/$Wi.
		Their functional form is a sum of two terms, one for each $u^{(i)}$, each term having the exact same form as the two time scales toy model predicts.
		In the particular case of $\mu^{ss}$, and up to some relabeling of the constants, it yields
		\begin{equation}
		\label{eqMuss1}
			\mu^{ss}(\mathcal{I}, \text{Wi}) = \frac{\mu_1}{1 + M/\text{Wi}} + \frac{\mu_2 - \mu_1}{1 + \mathcal{I}_0/\mathcal{I}}\,.
		\end{equation}
		In the Bagnold regime, Wi$\gg1$, so that Eq.~(\ref{eqMuss1}) reduces to the $\mu(\mathcal{I})$ law (\ref{eqMuI}).
		The comparison between Eq.~(\ref{eqMuss1}) and the numerical solution to the GITT equations (\ref{eqITT2}) is displayed in Fig.~\ref{FigMuss}.

		One last feature of the toy models is their ability to be easily generalized to the case of suspensions of granular particles in a simple liquid of
		viscosity $\eta_\infty$.
		In that case, a new time scale arises, associated with the microscopic motion of particles in a viscous fluid, given by $t_\eta = \eta_\infty/P$,
		leading to the definition of a new dimensionless number $\mathcal{J}= t_\gamma/t_\eta$.
		If $t_m\gg t_\eta$, which corresponds to suspensions in a liquid of very low viscosity, the cage effect plateau begins at $t_m$, independent of the
		value of $\eta_\infty$.
		In that case, $\mu^{ss}$ is still given by Eq.~(\ref{eqMuss1}).
		For higher viscosities however, whenever $t_\eta\gg t_m$, the plateau extends from $t_\eta$ to $t_\Gamma$ or $t_\gamma$ (depending on which one controls the
		decay), so that the rate $\Gamma^{(1)}$ in the two step decay ansatz (\ref{eqPhi2s}) depends on $t_\eta$ instead of $t_m$, and
		$u^{(1)}= t_\gamma/t_\eta$.
		As a result, $\mu^{ss}$ becomes
		\begin{equation}
		\label{eqMuss2}
			\mu^{ss}(\mathcal{J}, \text{Wi}) = \frac{\mu_1}{1 + M/\text{Wi}} + \frac{\mu_2 - \mu_1}{1 + \mathcal{J}_0/\mathcal{J}}\,,
		\end{equation}
		where it should be noted that, apart from $\mathcal{J}_0$, all constants are the same as for the same granular particles evolving without surrounding
		fluid according to Eq.~(\ref{eqMuss1}).
		This has been noted earlier in experiments on submarine granular flows \cite{Cassar05}.

\section{Application to extensional flows}

	\subsection{The stress tensor}

		In the simple shear flow, the only nonzero components of the flow tensor $\kappa$ are outside of its diagonal; however, this is not true in general.
		The incompressibility only imposes the nullity of its trace.
		In the following, we focus on a complementary family of flows where the only nonzero components of $\kappa$ are on its diagonal: the extensional flows.
		We study two examples, a 2D flow and a 3D flow.

		\subsubsection{Planar extension}

			The planar extensional flow corresponds to an elongation of the fluid in one direction --- hereafter the $x$ direction --- combined with a
			contraction in one of the orthogonal directions --- hereafter the $y$ direction.
			Its flow tensor can be written
			\begin{equation}
			\label{eqkpe}
				\kappa^{pe} = \left(
				\begin{array}{ccc}
					\dot\gamma & 0           & 0 \\
					0          & -\dot\gamma & 0 \\
					0          & 0           & 0
				\end{array}\right) \,.
			\end{equation}

			From Eq.~(\ref{eqgeff}), we can estimate its effective shear rate:
			\begin{equation}
			\label{eqgepe}
				\dot{\gamma}^{eff}_{pe} = 2\dot\gamma \,.
			\end{equation}
			The shear stress, defined in Eq.~(\ref{eqs0}) is given in that case by $\sigma_0^{pe}= \left<\sigma_{xx}\right> - \left<\sigma_{yy}\right>$;
			note in particular that it does not involve any component of $\sigma_{\alpha\beta}$ outside of the diagonal.
			Thus, in order to get the evolution of the rheological observables from Eq.~(\ref{eqITT2}), we need the three following combinations of vertex
			integrals:
			\begin{equation}
			\label{eqJpe}
				\begin{split}
					& \mathbb{J}_{xx}^{\quad xx} - \mathbb{J}_{xx}^{\quad yy} = \frac{1+\varepsilon}{2S_k^2}\frac{k^2}{k(-t)}
					\bigg[\frac{\sigma_\perp \Delta\sigma}{3\pi^2}\big(\dot\gamma t\big) \\
					& \quad + \frac{k\Delta \sigma^2}{30\pi^2}\Big(1+4\dot\gamma t+\dot\gamma^2 t^2\Big) \bigg] \\[0.2cm]
					& \mathbb{J}_{yy}^{\quad xx} - \mathbb{J}_{yy}^{\quad yy} = \frac{1+\varepsilon}{2S_k^2} \frac{k^2}{k(-t)}
					\bigg[\frac{\sigma_\perp \Delta\sigma}{3\pi^2}\big(\dot\gamma t\big) \\
					& \quad - \frac{k\Delta \sigma^2}{30\pi^2}\Big(1-4\dot\gamma t+\dot\gamma^2 t^2\Big) \bigg] \\[0.2cm]
					& \mathbb{J}_{zz}^{\quad xx} - \mathbb{J}_{zz}^{\quad yy} = \frac{1+\varepsilon}{2S_k^2}\frac{k^2}{k(-t)}
					\bigg[\frac{\sigma_\perp \Delta\sigma}{3\pi^2}\big(\dot\gamma t\big) \\
					& \quad + \frac{k\Delta \sigma^2}{15\pi^2}\big(\dot\gamma t\big) \bigg] \,,
				\end{split}
			\end{equation}
			Note that each contribution consists of two terms: one term proportional to $\sigma_\perp\Delta\sigma$ which is the same along any direction,
			and therefore only contributes to the pressure, and one term proportional to $\Delta\sigma^2$.
			By combining the vertices Eq.~(\ref{eqJpe}) and the GITT equations (\ref{eqITT2}), the pressure, shear stress, and effective friction coefficient
			can be computed numerically.

			In order to get a better understanding of the structure of the rheology in the planar extensional flow, let us use our toy models to reexpress the equations above.
			Since the combinations of $\sigma_\perp$ and $\Delta\sigma$, and therefore of structure factors and other $k$-dependent quantities, are the same
			as in the case of the simple shear, we can still use the constants $S_0$ and $S_1$ used in Eq.~(\ref{eqs0ss0}) and Eq.~(\ref{eqPss0}) to express
			our results, what will make the comparisons easier.

			We finally get for the shear stress and the overpressure:
			\begin{equation}
			\label{eqrhpe}
				\begin{split}
					& \sigma_0^{pe} = 4S_1\big(\mathcal{K}_0 + \mathcal{K}_2 \big)\\
					& \Delta P^{pe} = 4\left(S_0 + \frac{5}{3} S_1\right) \mathcal{K}_1\,.
				\end{split}
			\end{equation}
			Although for planar extensional flows the overpressure has a very similar structure to that of the simple shear flow (\ref{eqPss0}), the shear stress
			gets an additional term proportional to $\mathcal{K}_2$.

			From Eq.~(\ref{eqrhpe}), we deduce that $\Delta P^{pe} = 4 \Delta P^{ss}$, and that $\sigma_0^{pe}$ now takes the form
			\begin{equation}
			\label{eqs0pe}
				\sigma_0^{pe} = \frac{\sigma_y^{pe}}{1 + \overline{\gamma}_c\,u}\left[1 + \frac{\overline{\gamma}_c^2}{2(1 + \overline{\gamma}_c\,u)^2}\right]\,,
			\end{equation}
			where $\sigma_y^{pe} =4 \sigma_y^{ss}$ and $u = \Gamma/\dot{\gamma}^{eff}$.

			\begin{widetext}

			\begin{figure}
				\begin{center}
					\includegraphics[scale=0.6]{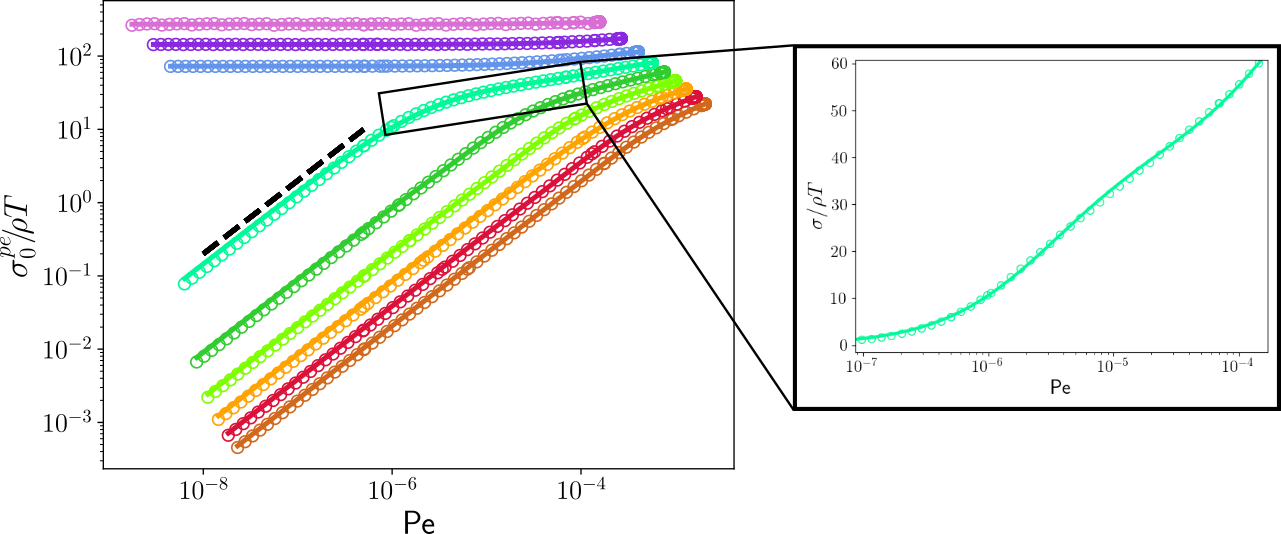}
				\end{center}
				\caption{Evolution of the dimensionless shear stress in the planar extension flow with the P\'eclet number for various packing fractions from
				$\varphi=0.42$ (bottom) to $\varphi = 0.58$ (top).
				The open circles are the numerical data gotten from the full GITT equation (\ref{eqITT2}); the full lines are curve fitting with help
				of the toy model expressions (\ref{eqrhpe}).
				The dashed line indicates the $1/$Pe slope for reference.
				The insert shows in greater detail the curve $\varphi=0.52$.}
				\label{FigSpe}
			\end{figure}

			\end{widetext}

			The overall behavior of $\sigma_0^{pe}$ is not modified much by the presence of the $\mathcal{K}_2$ term.
			Indeed, it is subdominant in the Newtonian regime, which can be verified numerically (see Fig.~\ref{FigSpe}).
			In the yielding regime, however, the system behaves as a yielding fluid with an effective yield stress
			$\sigma_y^{eff} = \sigma_y^{pe}(1+\overline{\gamma}_c^2/2)$.

\begin{figure}
				\begin{center}
					\includegraphics[scale=0.5]{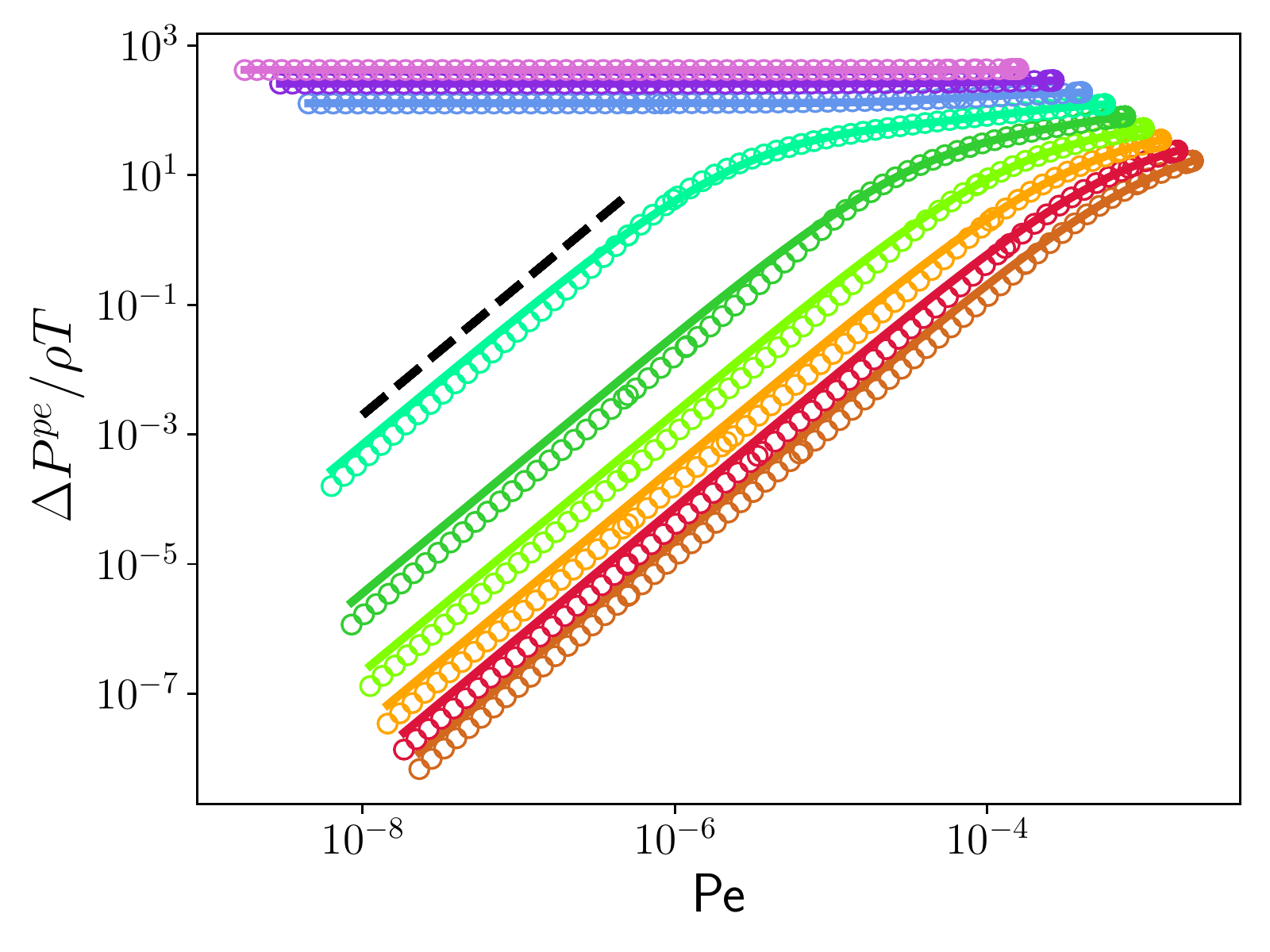}
				\end{center}
				\caption{Evolution of the dimensionless correction to the pressure in the planar extension flow with the P\'eclet number
				for various packing fractions from
				$\varphi=0.42$ (bottom) to $\varphi = 0.58$ (top).
				The open circles are the numerical data gotten from the full GITT equation (\ref{eqITT2}); the full lines are curve fitting with help
				of the toy model expressions (\ref{eqrhpe}).
				The dashed line indicates the $1/$Pe$^2$ slope for reference.}
				\label{FigPpe}
			\end{figure}

			The performance of the toy model can be assessed by fitting the numerical data obtained from the resolution of the GITT equation (\ref{eqITT2})
			with the toy model expressions of Eq.~(\ref{eqrhpe}).
			The results are shown in Figs.~\ref{FigSpe} and \ref{FigPpe}.
			The simple toy model agrees well with the numerical data, but it is necessary to upgrade Eq.~(\ref{eqrhpe}) to their three-time scale equivalent in
			order to be able to describe the data close to the MCT ideal granular glass transition that takes place around $\varphi=0.53$, and separates the
			yielding and Newtonian regimes at low Pe.
			Indeed, the distinction between Wi and Pe is necessary to reproduce subleading variations, such as the slowing of the growth of $\sigma_0^{pe}$
			around Pe$=10^{-5}$ for $\varphi=0.52$ displayed in the insert of Fig.~\ref{FigSpe}.

		\subsubsection{Uniaxial extension}

			A similar study can be performed on the uniaxial extension flow, which corresponds to a flow where the liquid is elongated along one direction
			--- $x$ in the following --- and contracted along both orthogonal directions.
			Its flow matrix is
			\begin{equation}
			\label{eqkue}
				\kappa^{ue} = \left(
				\begin{array}{ccc}
					\dot\gamma & 0             & 0             \\
					0          & -\dot\gamma/2 & 0             \\
					0          & 0             & -\dot\gamma/2
				\end{array}\right) \,.
			\end{equation}
			It is therefore a 3D flow.
			Its effective shear rate is related to $\dot\gamma$ by:
			\begin{equation}
			\label{eqgeue}
				\dot{\gamma}^{eff}_{ue} = \sqrt{3}\,\dot\gamma \,.
			\end{equation}

\begin{figure}
				\begin{center}
					\includegraphics[scale=0.5]{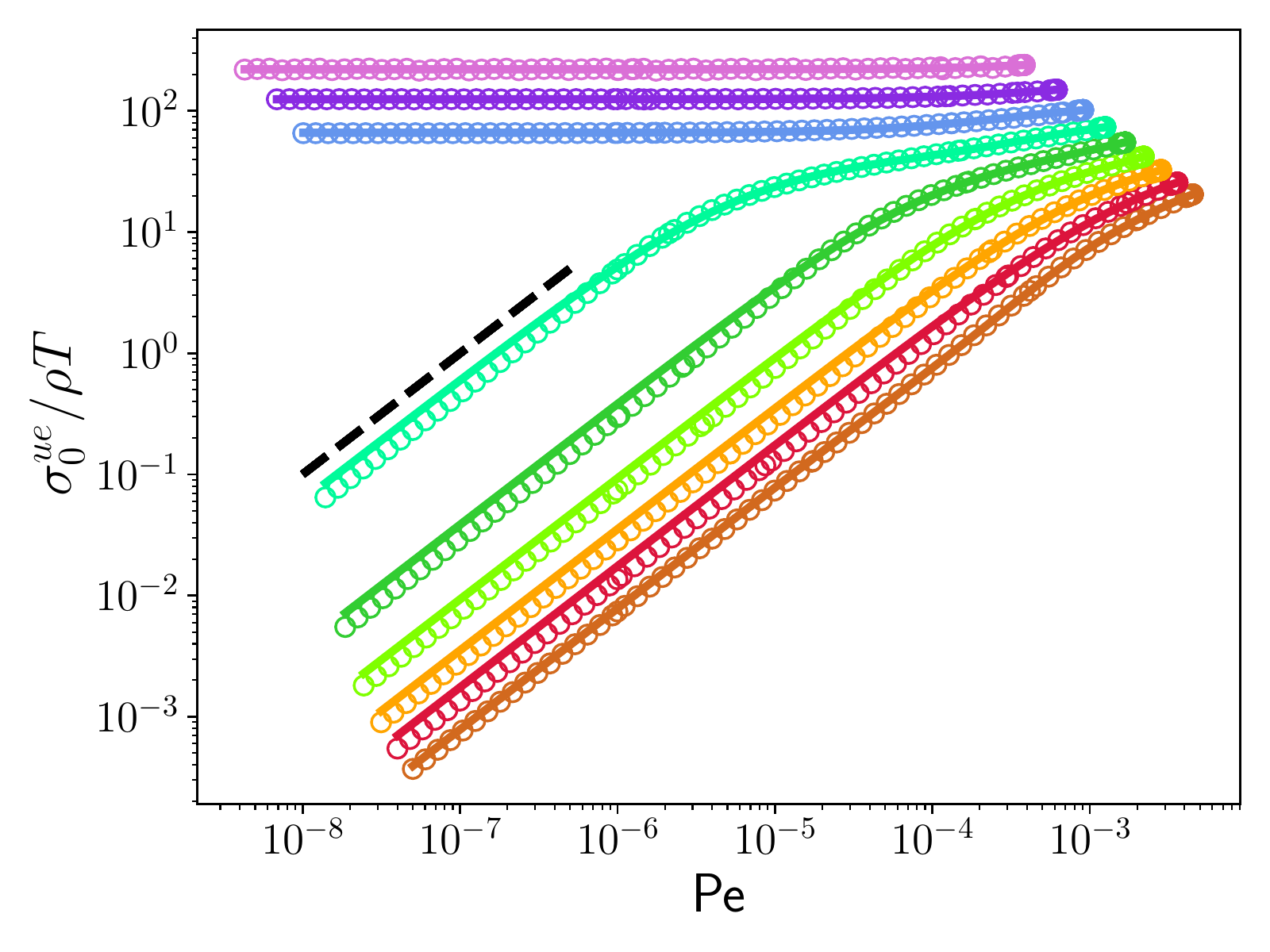}
				\end{center}
				\caption{Evolution of the dimensionless shear stress in the uniaxial extension flow with the P\'eclet number for various packing fractions from
				$\varphi=0.42$ (bottom) to $\varphi = 0.58$ (top).
				The open circles are the numerical data gotten from the full GITT equation (\ref{eqITT2}); the full lines are curve fitting with help
				of the toy model expressions (\ref{eqrhue}).
				The dashed line indicates the $1/$Pe slope for reference.}
				\label{FigSue}
			\end{figure}

\begin{figure}
				\begin{center}
					\includegraphics[scale=0.5]{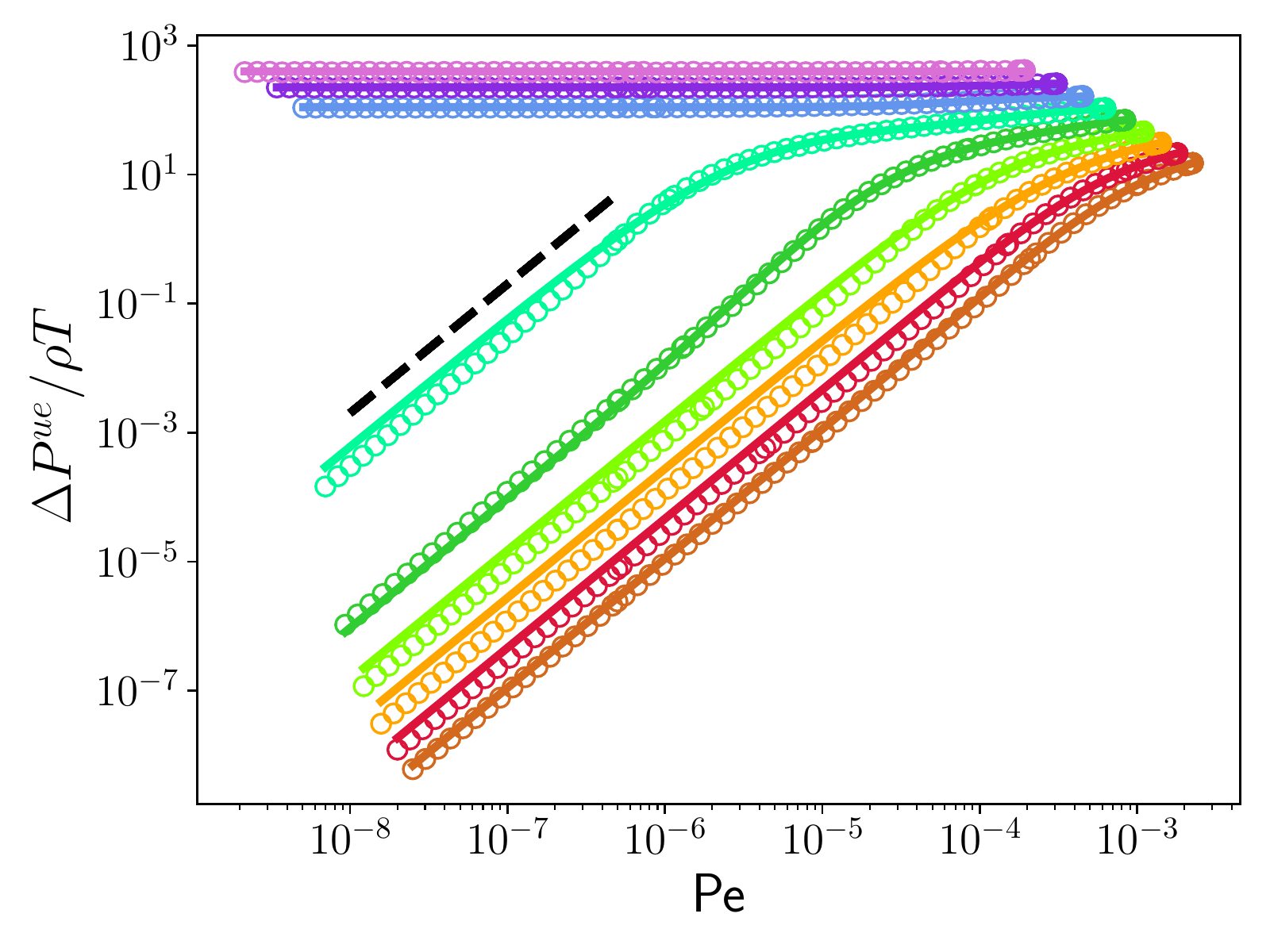}
				\end{center}
				\caption{Evolution of the dimensionless correction to the pressure in the uniaxial extension flow with the P\'eclet number for various packing fractions from
				$\varphi=0.42$ (bottom) to $\varphi = 0.58$ (top).
				The open circles are the numerical data gotten from the full GITT equation (\ref{eqITT2}); the full lines are curve fitting with help
				of the toy model expressions (\ref{eqrhue}).
				The dashed line indicates the $1/$Pe$^2$ slope for reference.}
				\label{FigPue}
			\end{figure}

			The shear stress $\sigma_0 = \left<\kappa:\sigma\right>$ hence involves only diagonal components of the stress tensor.
			In order to compute $\sigma_0$ and $P$, we thus need the following combinations of vertex integrals:
			\begin{equation}
			\label{eqJue}
				\begin{split}
					& \mathbb{J}_{xx}^{\quad xx} - \frac{\mathbb{J}_{xx}^{\quad yy}}{2} - \frac{\mathbb{J}_{xx}^{\quad zz}}{2}
					= \frac{1+\varepsilon}{2S_k^2} \frac{k^2}{k(-t)} \\
					& \quad\times\bigg[\frac{\sigma_\perp \Delta\sigma}{4\pi^2}\big(\dot\gamma t\big)\big(1+\dot\gamma t/4\big) \\
					& \quad +\frac{k\Delta \sigma^2}{60\pi^2}\Big(2+7\dot\gamma t+11\dot\gamma^2 t^2/4\Big) \bigg] \\[0.2cm]
					& \mathbb{J}_{yy}^{\quad xx} - \frac{\mathbb{J}_{yy}^{\quad yy}}{2} - \frac{\mathbb{J}_{yy}^{\quad zz}}{2}
					= \frac{1+\varepsilon}{2S_k^2} \frac{k^2}{k(-t)} \\
					& \quad\times\bigg[\frac{\sigma_\perp \Delta\sigma}{4\pi^2}\big(\dot\gamma t\big)\big(1+\dot\gamma t/4\big) \\
					& \quad - \frac{k\Delta \sigma^2}{60\pi^2}\Big(1-4\dot\gamma t - \dot\gamma^2 t^2/2 \Big) \bigg] \\[0.2cm]
					& \mathbb{J}_{zz}^{\quad xx} - \frac{\mathbb{J}_{zz}^{\quad yy}}{2} - \frac{\mathbb{J}_{zz}^{\quad zz}}{2}
					= \frac{1+\varepsilon}{2S_k^2} \frac{k^2}{k(-t)} \\
					& \quad\times\bigg[\frac{\sigma_\perp \Delta\sigma}{4\pi^2}\big(\dot\gamma t\big)\big(1+\dot\gamma t/4\big) \\
					& \quad - \frac{k\Delta \sigma^2}{60\pi^2}\Big(1 - 4\dot\gamma t- \dot\gamma^2 t^2/2 \Big) \bigg] \,.
				\end{split}
			\end{equation}
			These expressions are sufficient to compute the pressure and shear stress numerically.
			The results are displayed in Figs.~\ref{FigSue} and \ref{FigPue}.

			Let us now analyze the structure of the stress tensor through the two time scales toy model.
			First, the vertices in Eq.~(\ref{eqJue}) are combinations of $\sigma_\perp\Delta\sigma$ and $\Delta\sigma^2$, they can therefore be expressed in 
			terms of $S_0$ and $S_1$ only.
			The decomposition of $\sigma_0$ and $\Delta P$ on the $\mathcal{K}$ integrals writes
			\begin{equation}
			\label{eqrhue}
				\begin{split}
					& \sigma_0^{ue} = 3S_1 \big(\mathcal{K}_0 + \mathcal{K}_1 + 3\mathcal{K}_2/4 \big)\\[0.2cm]
					& \Delta P^{ue} = \big(3S_0 + 5 S_1 \big)\big( \mathcal{K}_1 + \mathcal{K}_2/4 \big)\,.
				\end{split}
			\end{equation}
			The structure of $\sigma_0$ is thus further enriched by the presence of $\mathcal{K}_1$, and in contrast to the simple shear and
			planar extension flows, $\Delta P$ is not merely proportional to $\mathcal{K}_1$ anymore.
			This richer structure is not striking on the numerical data in Figs.~\ref{FigSue} and \ref{FigPue}, because the role of the additional integrals is
			only subleading as discussed in the case of the planar extensional flow.
			However, as we discuss below, it has a measurable impact on the behavior of the effective friction coefficient.

			The final toy model expressions for the uniaxial extension flow are:
			\begin{equation}
			\label{eqrhue1}
				\begin{split}
					& \sigma_0^{ue} = \frac{\sigma_y^{ue}}{1 + \overline{\gamma}_c\,u}\left[1 
					+ \frac{\overline{\gamma}_c}{2(1 + \overline{\gamma}_c\,u)} + \frac{3\overline{\gamma}_c^2}{8(1 + \overline{\gamma}_c\,u)^2}\right]\\[0.2cm]
					& \Delta P^{ue} = \frac{P_1^{ue}}{\big(1 + \overline{\gamma}_c\,u)^2}\left[1 
					+ \frac{\overline{\gamma}_c}{4(1 + \overline{\gamma}_c\,u)}\right]\,,
				\end{split}
			\end{equation}
			where $\sigma_y^{ue} = 3\sigma_0^{ss}$ and $P_1^{ue} = 3 P_1^{ss}$.
			Note also that despite the increased complexity of the expressions (\ref{eqrhue1}) compared to Eq.~(\ref{eqs0ss1}) and (\ref{eqPss1}) they involve the
			same number of constants.
			The effective yield stress in the yielding regime is $\sigma_y^{ue}(1 + \overline{\gamma}_c/2 +3\overline{\gamma}_c/8)$, and the overpressure
			is also distorted to $P_1^{ue}(1+\overline{\gamma_c}/2)$ in this regime.

	\subsection{The effective friction coefficient}

		\subsubsection{General expressions}

\begin{figure}
				\begin{center}
					\includegraphics[scale=0.5]{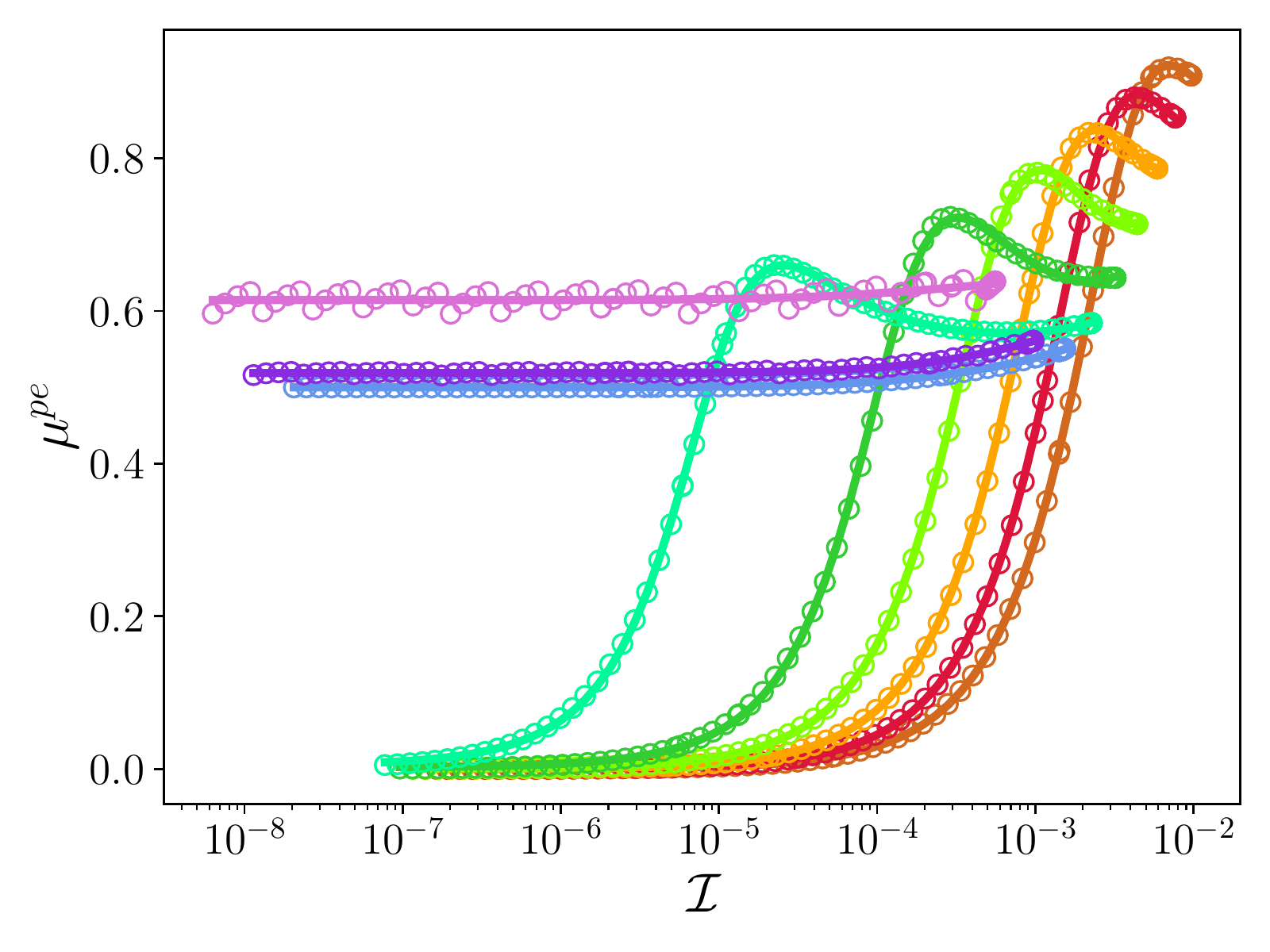}
				\end{center}
				\caption{Evolution of the effective friction coefficient in the planar extension flow with the inertial number for various packing fractions from
				$\varphi=0.42$ (bottom) to $\varphi = 0.58$ (top).
				The open circles are the numerical data gotten from the full GITT equation (\ref{eqITT2}); the full lines are curve fitting with help
				of the toy model expressions (\ref{eqmupe}).}
				\label{FigMupe}
			\end{figure}

\begin{figure}
				\begin{center}
					\includegraphics[scale=0.5]{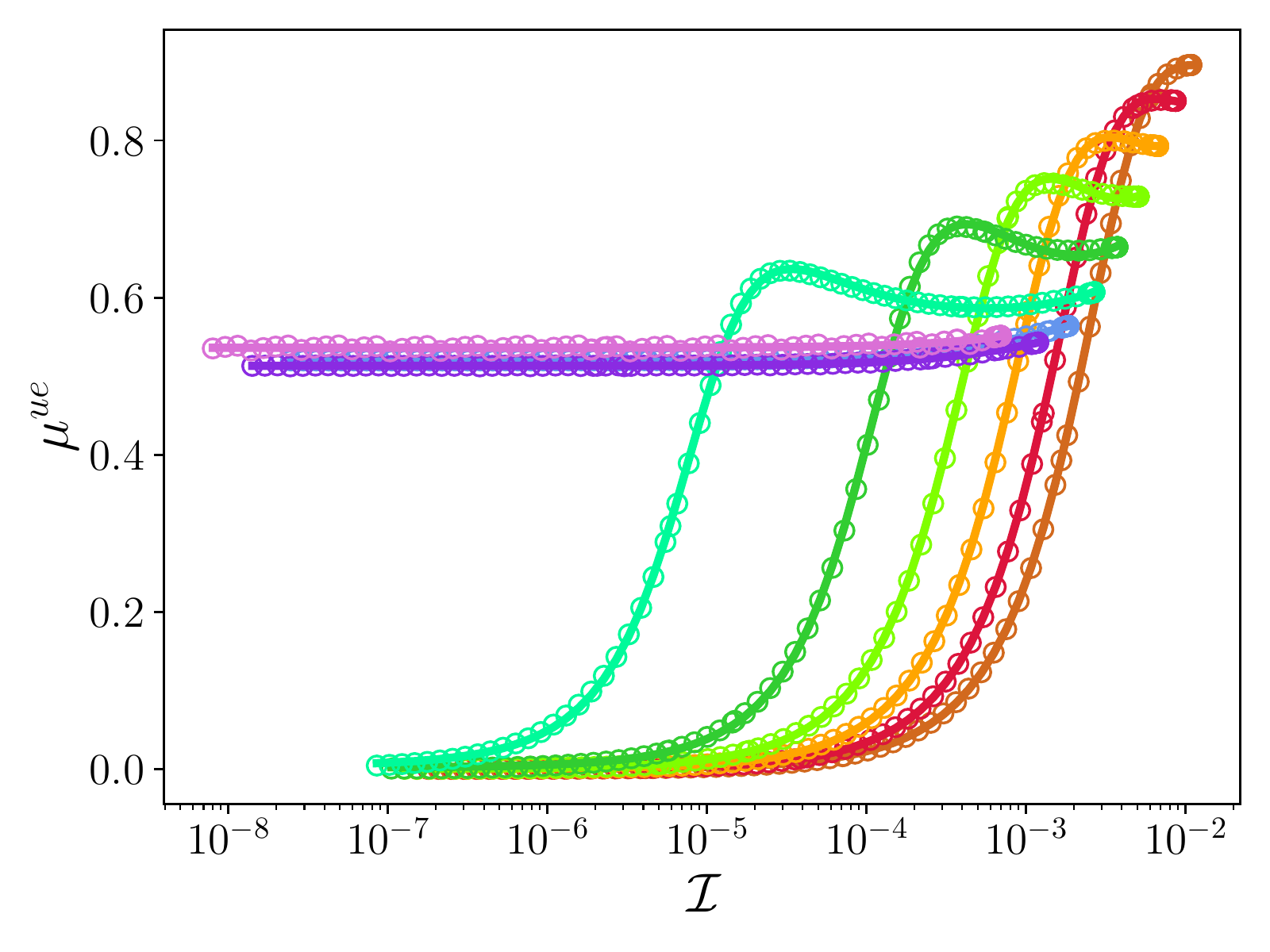}
				\end{center}
				\caption{Evolution of the effective friction coefficient in the uniaxial extension flow with the inertial number for various packing fractions from
				$\varphi=0.42$ (bottom) to $\varphi = 0.58$ (top).
				The open circles are the numerical data gotten from the full GITT equation (\ref{eqITT2}); the full lines are curve fitting with help
				of the toy model expressions (\ref{eqmuue}).}
				\label{FigMuue}
			\end{figure}

			The numerical expression of $\mu^{pe}$ and $\mu^{ue}$ can be directly evaluated from Eq.~(\ref{eqJpe}) and (\ref{eqJue}).
			The results are displayed on Figs.~\ref{FigMupe} and \ref{FigMuue}, respectively.
			These curves show two particular features that distinguish them from the case of simple shear flows: the maximum value of $\mu$ across the
			P\'eclet range is no longer a growing function of the packing fraction, and $\mu$ is no longer a monotonous function of Pe.

			Let us try to explain this with the two-time scales toy model.
			The effective friction coefficient can be derived directly from Eq.~(\ref{eqrhpe}) for the planar extension flow:
			\begin{equation}
			\label{eqmupe0}
				\mu^{pe} = \frac{4S_1\big(\mathcal{K}_0 + \mathcal{K}_2\big)}{P_0 + 4\big(S_0 + 5/3S_1\big)\mathcal{K}_1}\,,
			\end{equation}
			and Eq.~(\ref{eqrhue}) for the uniaxial extension:
			\begin{equation}
			\label{eqmuue0}
				\mu^{ue} = \frac{3S_1\big(\mathcal{K}_0 + \mathcal{K}_1 +3\mathcal{K}_2/4\big)}
				{P_0 + \big(3S_0 + 5S_1\big)\big(\mathcal{K}_1 + \mathcal{K}_2\big)}\,.
			\end{equation}

			Hence, in both cases, $\mu$ can be expressed as a Pad\'e $P[2/3]$, but with only four independent coefficients.
			More precisely, in the planar extension case, $\mu$ can be written as
			\begin{equation}
			\label{eqmupe}
				\mu^{pe} = \frac{M_{00}^{pe} + M_{01}^{pe}\, u + M_{02}^{pe}\, u^2}{1 + M_{11}^{pe}\, u + M_{12}^{pe}\, u^2 + M_{13}^{pe}\, u^3}\,,
			\end{equation}
			with coefficients $M_{00}^{pe} = \mu_y^{pe}(1+\overline{\gamma}_c^2/2)$, $M_{01}^{pe} = 2\overline{\gamma}_c\mu_y^{pe}$,
			$M_{02}^{pe} = \overline{\gamma}_c^2\mu_y^{pe}$, $M_{11}^{pe} = (3P_0 + P_1^{pe})\overline{\gamma}_c/P_y^{pe}$,
			$M_{12}^{pe} = 3\overline{\gamma}_c^2P_0/P_y^{pe}$ and $M_{13}^{pe} = \overline{\gamma}_c^3P_0/P_y^{pe}$
			expressed in terms of the pressure in the yielding regime $P_y^{pe} = P_0 + P_1^{pe}$ and the characteristic friction coefficient
			$\mu_y^{pe} = \sigma_y^{pe}/P_y^{pe}$.
			Similarly, in the uniaxial extension case, it can be written
			\begin{equation}
			\label{eqmuue}
				\mu^{ue} = \frac{M_{00}^{ue} + M_{01}^{ue}\, u + M_{02}^{ue}\, u^2}{1 + M_{11}^{ue}\, u + M_{12}^{ue}\, u^2 + M_{13}^{ue}\, u^3}\,,
			\end{equation}
			with coefficients $M_{00}^{ue} = \mu_y^{ue}(1+\overline{\gamma}_c/2 + 3\overline{\gamma_c}^2/8)$,
			$M_{01}^{ue} =\overline{\gamma}_c(2+\overline{\gamma}_c/2)\mu_y^{ue}$,
			$M_{02}^{ue} = \overline{\gamma}_c^2\mu_y^{ue}$, $M_{11}^{ue} = (3P_0 + P_1^{ue})\overline{\gamma}_c/P_y^{ue}$,
			$M_{12}^{ue} = 3\overline{\gamma}_c^2P_0/P_y^{ue}$ and $M_{13}^{ue} = \overline{\gamma}_c^3P_0/P_y^{ue}$
			expressed in terms of the pressure in the yielding regime $P_y^{ue} = P_0 + P_1^{ue}(1 + \overline{\gamma}_c/4)$, now explicitly dependent on $\overline{\gamma}_c$,
			and the characteristic friction coefficient $\mu_y^{ue} = \sigma_y^{ue}/P_y^{ue}$.

			Thus, even though the richer structure of $\sigma_0$ and $P$ in terms of powers of $u$ leads to very minor changes on the evolution of those two
			quantities, taken separately, compared to the simple shear case, its impact on the effective friction coefficient has a measurable effect.
			Indeed, the ansatz (\ref{eqMuss0}) can only lead to a monotonous evolution of $\mu$ with the shear rate, contradicting the
			evolutions observed in the numerical data in Figs.~\ref{FigMupe} and \ref{FigMuue}.
			In that respect, we can say that the rheology of extensional flows is qualitatively different from that of simple shear flows.

			Finally, extending the above to the three time scales toy model (when we recall that $\mu$ becomes the sum of two terms, both with a functional form
			similar to that of the expressions (\ref{eqmupe}) and (\ref{eqmuue}) but expressed in terms of $u^{(1)}\propto1/$Pe and $u^{(2)}\propto1/$Wi)
			provides expressions that fit the numerical data with a satisfactory precision, as can be seen in Figs.~\ref{FigMupe} and \ref{FigMuue}.
			Note that this model generalizes trivially to granular suspensions, with an effective friction coefficient given by the formulas above
			for $t_\eta\ll t_m$, and by the same functional form of the three time scales toy model with $u^{(1)}=t_\gamma/t_\eta\propto1/\mathcal{J}$ and $u^{(2)}\propto1/$Wi
			if $t_\eta\gg t_m$.

		\subsubsection{The Bagnold regime}

			Let us now discuss the case of the Bagnold regime, where most dry granular flows lie, and where the $\mu(\mathcal{I})$-law holds for simple shear flows.
			Since these effects are subleading, compared to the effects discussed above, the three time scales version of the toy model is required.
			In that case, the effective friction coefficient is decomposed as $\mu = \mu^{(1)} + \mu^{(2)}$, where $\mu^{(1)}$ and $\mu^{(2)}$ have the functionals form
			of Eqs.~(\ref{eqmupe}) and (\ref{eqmuue}) with appropriate sets of constants.

			Let us first focus on $\mu^{(2)}$. It is defined as a function of $u^{(2)} = t_\gamma/t_\Gamma\propto1/$Wi.
			In the Bagnold regime, the shear rate is very strong at the scale of the system evolution, namely the decay of $\Phi(t)$ is caused
			by advection, and not structural relaxations.
			This means that Wi$\gg1$, so that $u^{(2)}\ll1$, hence, by use of Eq.~(\ref{eqmupe}), $\mu^{(2)}\simeq M_{00}^{pe,(2)}$ which is a constant.
			As for $\mu^{(1)}$, it is a function of $u^{(1)}\propto1/$Pe $\propto 1/\mathcal{I}$, which therefore contains the part of $\mu^{pe}$ that varies
			with the inertial number.
			It can also be checked from Eq.~(\ref{eqmupe}) that when $\mathcal{I}\rightarrow0$, $\mu^{(1)}\rightarrow 0$.
			By analogy with the $\mu(\mathcal{I})$-law (\ref{eqMuI}), we can identify the constant value of $\mu^{(2)}$ with the $\mathcal{I}\rightarrow0$ limit
			of $\mu^{pe}$, and call it $\mu_1$.
			All in all, dropping the $(1)$ superscript in the constants of the $\mu^{(1)}$ term, the corresponding evolution of $\mu^{pe}$ with $\mathcal{I}$ writes
			\begin{equation}
			\label{eqmupeI}
				\mu^{pe} = \mu_1 + \frac{M_{00}^{pe} + M_{01}^{pe}/\mathcal{I} + M_{02}^{pe}/\mathcal{I}^2}
				{1 + M_{11}^{pe}/\mathcal{I} + M_{12}^{pe}/\mathcal{I}^2 + M_{13}^{pe}/\mathcal{I}^3}\,.
			\end{equation}
			Equation (\ref{eqmupeI}) is the equivalent to the $\mu(\mathcal{I})$-law in the case of planar extensional flows as predicted by the three
			time scales toy model.
			It is equivalent to a $P[3/3]$ Pad\'e approximant.

			\begin{figure}
				\begin{center}
					\includegraphics[scale=0.5]{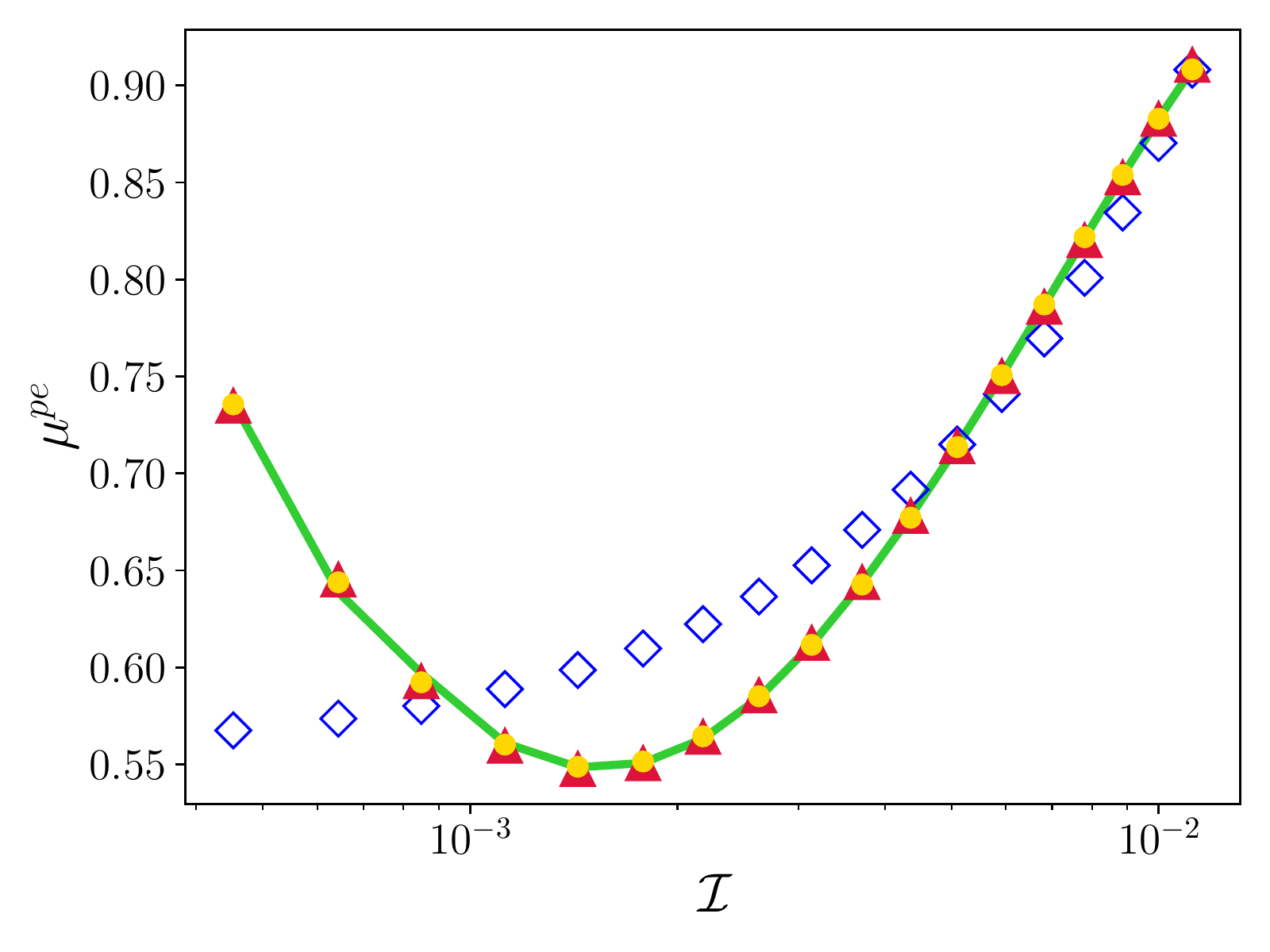}
				\end{center}
				\caption{Evolution of the effective friction coefficient in the planar extension flow as a function of the inertial number.
				The full line is the numerical data resulting from the resolution of the GITT equations of Eq.~(\ref{eqITT2}); the circles are the best fit with a
				Pad\'e ansatz $P[3/3]$, the triangles correspond to $P[2/2]$ and the diamonds to $P[1/1]$.}
				\label{FigMuIpe}
			\end{figure}
			
			However, we can go further.
			In the simple shear case, the form equivalent to Eq.~(\ref{eqmupeI}) is not exactly the $\mu(\mathcal{I})$-law, but a more involved
			Pad\'e $P[2/2]$ approximant \cite{Coquand20g}.
			The excellent agreement between the simpler $\mu(\mathcal{I})$-law and the experimental or numerical data suggests that most of the physics can be captured
			by a simpler ansatz with fewer fitting coefficients, a $P[1/1]$ function in that case.
			
			We can look for a similar formula in our system as well.
			From the structure of the three time scales toy model in the high Weissenberg number regime, we know that $\mu$ can be written
			as a sum of a constant $\mu_1$ and a rational fraction.
			The possible structures are thus $P[3/3]$ --- corresponding to the full three time scales toy model of Eq.~(\ref{eqmupeI}) --- $P[2/2]$ and
			$P[1/1]$ --- corresponding to simpler ansatz that one can propose to reduce the number of constants in the equivalent of the $\mu(\mathcal{I})$-law.
			The results are displayed in Fig.~\ref{FigMuIpe}.
			We can see in this figure that, even if the $P[3/3]$ ansatz of Eq.~(\ref{eqmupeI}) performs well as expected, the $P[2/2]$ provides a very good
			description of the numerical results too.
			The $P[1/1]$ form of Eq.~(\ref{eqMuI}) on the other hand is not rich enough to account for the nonmonotonous evolution of $\mu^{pe}$ with $\mathcal{I}$.
			The usual form of the $\mu(\mathcal{I})$ must therefore be updated to be able to capture the phenomenology of planar extensional flows.

			Another way to understand the reduction to the $P[2/2]$ ansatz is the following: we established that in the two-time scales toy model (\ref{eqmupeI}) 
			$\mu^{pe}$ takes the form of a rational fraction with a polynomial of order three in the denominator, but with only four independent constants.
			In the case of the three time scales toy model in the Wi$\gg1$ regime, the $\mu^{(2)}$ contribution adds one independent constant, so there are five
			of them in total.
			This corresponds exactly to the number of independent coefficients of a $P[2/2]$ ansatz.

			\begin{figure}
				\begin{center}
					\includegraphics[scale=0.5]{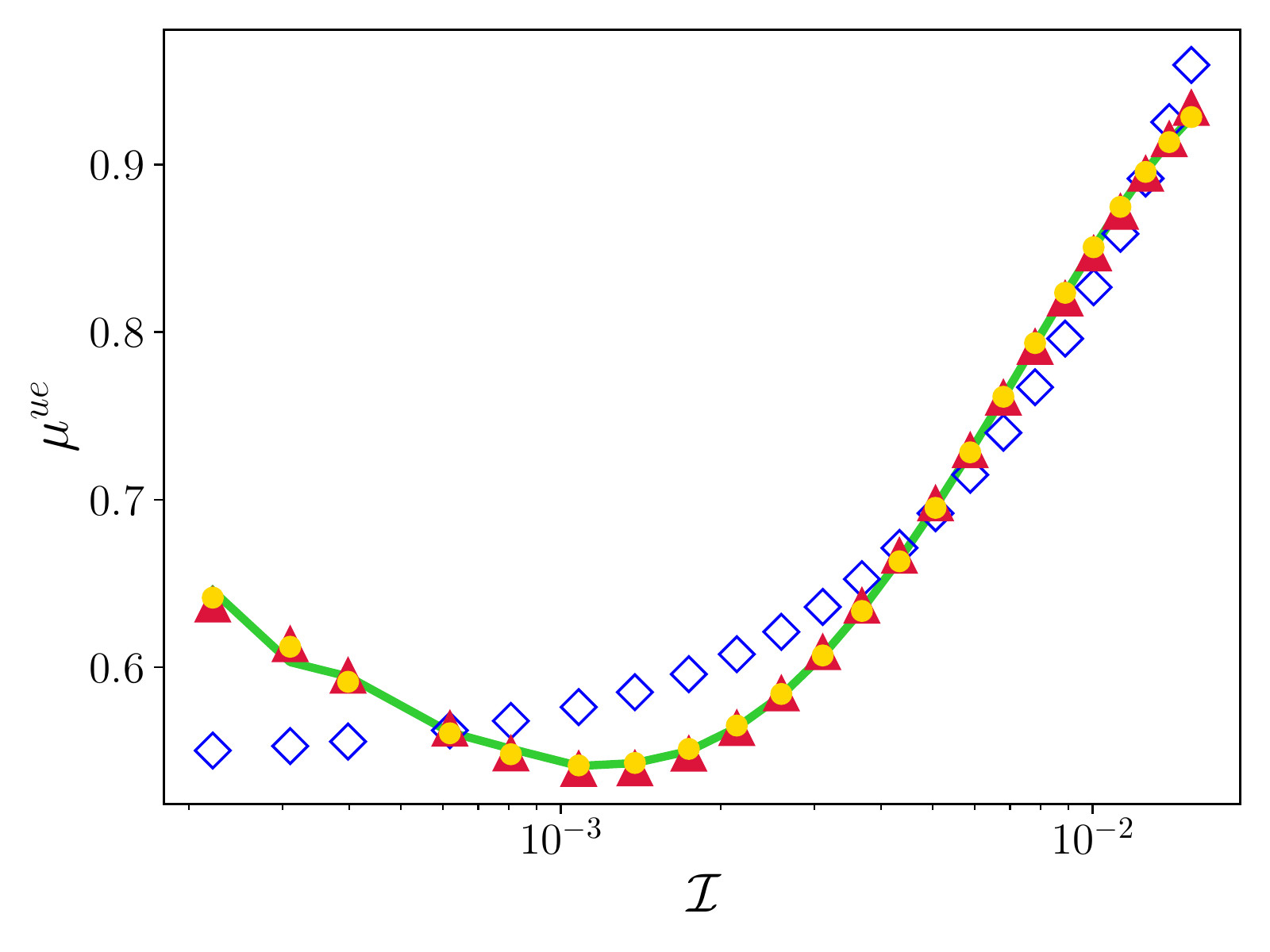}
				\end{center}
				\caption{Evolution of the effective friction coefficient in the uniaxial extension flow as a function of the inertial number.
				The full line is the numerical data resulting from the resolution of the GITT equations of Eq.~(\ref{eqITT2}); the circles are the best fit with a
				Pad\'e ansatz $P[3/3]$, the triangles correspond to $P[2/2]$ and the diamonds to $P[1/1]$.}
				\label{FigMuIue}
			\end{figure}

			A similar study can be conducted for uniaxial extensional flows.
			Since $\mu$ in Eqs.~(\ref{eqmuue}) and (\ref{eqmupe}) have similar forms, the reasoning in that case is exactly the same, only the constants are changed.
			Hence in the Bagnold regime, $\mu^{ue,(2)}\simeq \mu_1^{ue}$, which is a constant, and the three time scales toy model describes $\mu^{ue}(\mathcal{I})$
			as a $P[3/3]$ ansatz.
			We also compared different types of Pad\'e ansatz to get the minimal $\mu^{ue}(\mathcal{I})$ ansatz with reasonable performance.
			The results are displayed in Fig.~\ref{FigMuIue}.
			As in the case of planar extension, the evolution of $\mu^{ue}(\mathcal{I})$ is well captured by a $P[2/2]$ ansatz.
			The $P[1/1]$ used in the simple shear case, however, fails for the same reason, namely, its monotonous behavior.

			All in all, the $\mu(\mathcal{I})$ can be generalized as
			\begin{equation}
			\label{eqMuImod}
				\mu(\mathcal{I}) = \mu_1 + \frac{\mu_2 - \mu_1 + \mathcal{I}_1/\mathcal{I}}{1 + \mathcal{I}_0/\mathcal{I} + \mathcal{I}_2^2/\mathcal{I}^2}\,.
			\end{equation}
			This expression generalizes the well-known law (\ref{eqMuI}), and is valid in any flow configuration.
			In the case of pure shear deformations, the constants $\mathcal{I}_1$ and $\mathcal{I}_2$ can be safely neglected.

	\subsection{The Trouton ratio}

		Consider two flow configurations described respectively by the flow matrices $\kappa_a$ and $\kappa_b$.
		We want to compare the fluid's response to shear in both configurations.
		We already saw a way to compare both flow geometries, by the definition of an effective shear stress $\dot\gamma^{eff}$ in Eq.~(\ref{eqgeff}).
		However, in an experiment for example, it can be easier to get access to the real shear stress $\dot\gamma$, defined by the rate of deformation in a
		given direction, rather than the effective one.
		It is therefore useful to be able to read the deviatoric part of the stress tensor in two ways:
		\begin{equation}
		\label{eqeeff}
			\sigma:D = 2 \,\eta(\dot\gamma) \,\big(\dot\gamma^{eff}\big)^2 = 2 \,\eta^{eff}(\dot\gamma)\, \dot\gamma^2\,,
		\end{equation}
		where we recalled that in general, the viscosity of the fluid depends on the shear rate.
		The fluid's response to shear at a given shear rate can then characterised by its effective viscosity $\eta^{eff}$.

		Let us define $\mathcal{T}_a^b$ as the ratio of the effective shear viscosities of the flow configurations $a$ and $b$ (for sake of simplicity
		we get rid of the $eff$ superscript, a viscosity dependent on a flow geometry being implicitly an effective one):
		\begin{equation}
		\label{eqT0}
			\mathcal{T}_a^b = \frac{\eta_b}{\eta_a}\,.
		\end{equation}

		If the fluid is in the Newtonian regime, the shear viscosity $\eta$ is independent of the shear rate, so that the $\dot\gamma$ dependence of $\eta^{eff}$ in
		Eq.~(\ref{eqeeff}) is the same as that of $\big(\dot\gamma^{eff}\big)^2$.
		Consequently,
		\begin{equation}
		\label{eqT1}
			\mathcal{T}_a^b \underset{t_\Gamma \ll t_\gamma}{\longrightarrow} \frac{\big(\dot{\gamma}_b^{eff}\big)^2}{\big(\dot{\gamma}_a^{eff}\big)^2}\,,
		\end{equation}
		where $t_\Gamma\ll t_\gamma$ is a way to indicate that the Newtonian limit is taken.
		Hence, for any fluid, and in any two flow configurations, the limit of the ratio of the effective shear viscosities in the Newtonian regime is equal to the
		ratio of the squared effective shear rates.

\begin{figure}
			\begin{center}
				\includegraphics[scale=0.5]{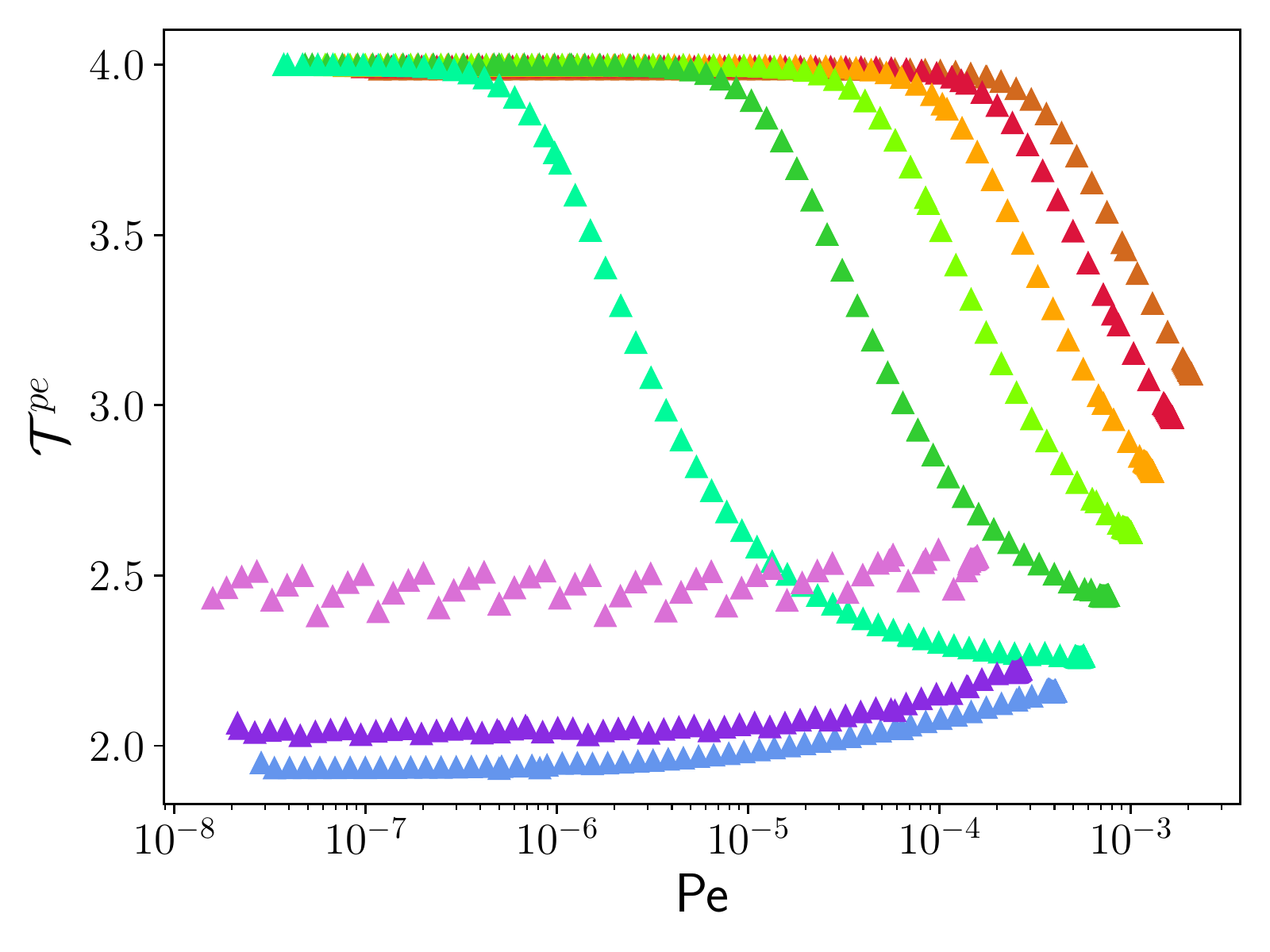}
			\end{center}
			\caption{Evolution of the Trouton ratio in the planar extension flow as a function of the P\'eclet number for various packing fractions between 
			$\varphi = 0.42$ (brown) and $\varphi = 0.58$ (purple) computed from the numerical resolution
			of the GITT equations (\ref{eqITT2}).}
			\label{FigTroutpe}
		\end{figure}

\begin{figure}
			\begin{center}
				\includegraphics[scale=0.5]{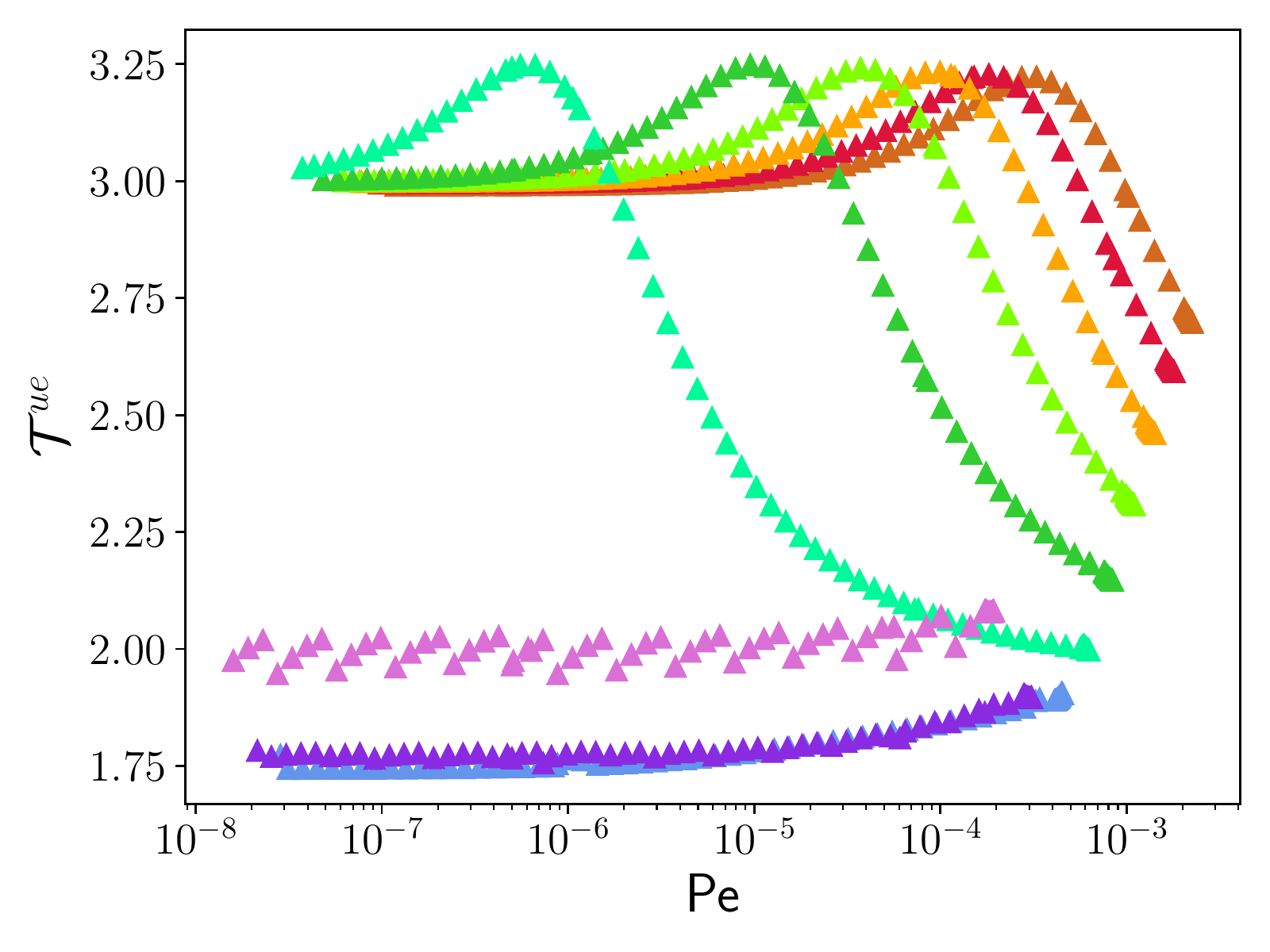}
			\end{center}
			\caption{Evolution of the Trouton ratio in the uniaxial extension flow as a function of the P\'eclet number for various packing fractions between 
			$\varphi = 0.42$ (brown) and $\varphi = 0.58$ (purple)
			computed from the numerical resolution of the GITT equations (\ref{eqITT2}).}
			\label{FigTroutue}
		\end{figure}

		In the particular case where the configuration $a$ is a simple shear flow, $\mathcal{T}$ is called the \textit{Trouton ratio} \cite{Trouton06}.
		Indeed, since in simple shear flows $\dot\gamma^{eff} = \dot\gamma$, they constitute a useful reference point.
		Our previous study allows us to discuss the evolution of the Trouton ratios of the planar and extensional flows, $\mathcal{T}^{pe}$ and $\mathcal{T}^{ue}$
		respectively, for granular liquids and suspensions.
		The results are displayed in Figs.~\ref{FigTroutpe} and \ref{FigTroutue}.

		In order to facilitate the interpretation of the numerical results, we will discuss them in the light of the two time scales toy model.
		First let us discuss the case of planar extension.
		The Trouton ratio in that case is given by the ratio of $\eta^{pe}$ to $\eta^{ss}$ that can be related to their respective shear stresses (\ref{eqs0ss1})
		and (\ref{eqs0pe}) by Eq.~(\ref{eqs0}), and the definition of the effective viscosity (\ref{eqeeff}).
		The effective shear stresses are related by Eq.~(\ref{eqgepe}).
		Since $\Gamma$ and $\overline{\gamma}_c$ are determined by the dynamics of the unsheared reference state, $\Gamma^{pe} = \Gamma^{ss}$ and $\overline{\gamma}_c^{pe}
		=\overline{\gamma}_c^{ss}$.
		As a result, $u^{pe} = \Gamma^{pe}/\dot{\gamma}^{eff}_{pe} = \Gamma/(2\dot\gamma) = u^{ss}/2$.
		Finally, we also established that $\sigma^{pe}_y = 4\sigma_y^{ss} = 4G_\infty\overline{\gamma}_c$.
		Combining all these equations leads to:
		\begin{equation}
		\label{eqTpe}
			\mathcal{T}^{pe} = 2\,\frac{1 + \overline{\gamma}_c\,u}{1 + \overline{\gamma}_c\,u/2}\left[1 + \frac{\overline{\gamma}_c^2}
			{2\big(1 + \overline{\gamma}_c\,u/2\big)^2}\right]\,,
		\end{equation}
		where $u = u^{ss} = \Gamma/\dot\gamma$.

		As expected, in the Newtonian limit $u\gg1$, $\mathcal{T}^{pe}\rightarrow 4 = \big(\dot{\gamma}^{eff}_{pe}\big)^2/\dot{\gamma}^2$, which is consistent
		with Eq.~(\ref{eqT1}).
		This is of course no coincidence.
		Indeed, the value of the Trouton ratio in the Newtonian limit can be inferred directly from the full GITT equation (\ref{eqITT2}) through the use of the
		viscosity tensor $\Lambda$: Let us consider the computation of the viscosity of the liquid in the Newtonian regime, and the Trouton ratio.
		The viscosity is defined from $\sigma:\kappa$, or equivalently $\kappa:\Lambda:\kappa$. 
		For the simple shear flow, $\kappa^{ss}$ is a nilpotent matrix. It can therefore be checked that $\kappa^{ss}:\Lambda:\kappa^{ss} = \mathcal{B}_X^\sigma \dot\gamma^2$.

		In the Newtonian limit, only the lowest $\dot\gamma$ dependence is kept.
		Thus, the $\dot\gamma$ dependence of $\Phi$ can be neglected, so that $\mathcal{B}_X^\sigma$ is independent from the flow geometry in that limit.
		For the same reason, in the Newtonian limit, the contributions of $\mathcal{B}_1^\sigma$ and $\mathcal{B}_2^\sigma$ can be neglected compared to that
		of $\mathcal{B}_X^\sigma$ for any type of flow.
		As a result, for any flow of flow tensor $\kappa$, the Newtonian contribution to the viscosity  can be written in full generality as $\mathcal{B}_X^\sigma \kappa:X:\kappa
		 = \mathcal{B}_X^\sigma D:D/2 = \mathcal{B}_X^\sigma \big(\dot\gamma^{eff}\big)^2$.
		Since $\mathcal{B}_X^\sigma$ has the same value for the simple shear flow and for the more general flow,
		it follows that $\mathcal{T}=\big(\dot\gamma^{eff}\big)^2/\dot\gamma^2$ in the Newtonian regime.

		Pay attention to the fact that strictly speaking $\mathcal{B}_X^\sigma \kappa:X:\kappa$ is not a viscosity but a shear stress,
		an overall factor $\dot\gamma^{eff}$ should be added according to Eq.~(\ref{eqs0}).
		This factor is canceled in the Newtonian regime because $\mathcal{B}_X^\sigma
		\propto 1/u\propto \dot\gamma^{eff}$.

		Outside of the Newtonian regime, $\mathcal{T}^{pe}$ becomes non universal.
		A few of its properties can be understood from Eq.~(\ref{eqTpe}).
		For example, the limit of $\mathcal{T}^{pe}$ in the yielding regime is
		\begin{equation}
		\label{eqTpe2}
			\mathcal{T}^{pe}\underset{t_\Gamma \gg t_\gamma}{\longrightarrow}2\big(1+\overline{\gamma}_c^2/2\big)\,,
		\end{equation}
		which can be compared to Fig.~\ref{FigTroutpe} where $\mathcal{T}^{pe}\geqslant2$, and its saturation in the yielding regime is a growing
		function of the packing	fraction, and so does $\overline{\gamma}_c$ (remember that it is a typical strain scale).
		Some data points on Fig.~\ref{FigTroutpe} have $\mathcal{T}^{pe}\leqslant2$.
		However, we should keep in mind that deep into the yielding regime, $\eta\propto
		1/\dot\gamma$ is very big, so that the numerical precision of $\mathcal{T}^{pe}$,
		a ratio of two big numbers, may result in a slight underestimation of the Trouton ratio.
		The form of $\mathcal{T}^{pe}$ in Eq.~(\ref{eqTpe2}) is given by that of the effective yield stress of the planar extension flow defined above.
		Note, however, that the increase of $\mathcal{T}^{pe}$ with Pe is not captured at the level of the two time scales toy model; such subleading
		behaviors require the distinction between Wi and Pe.
		Interestingly, the evolution of $\mathcal{T}^{pe}$ in Fig.~\ref{FigTroutpe} is similar to that predicted by MCT for colloidal suspensions \cite{Brader09},
		thereby highlighting connections between the rheological behavior of both systems in stationary flows.
		The fact that $\mathcal{T}^{pe}$ and $\mathcal{T}^{ue}$ sharply decrease when the packing fraction is decreased is consistent with previous numerical studies
		on extensional flows of granular matter \cite{Cheal18}.

		A similar analysis can be conducted for the uniaxial elongation flow.
		In that case, the Trouton ratio can be written:
		\begin{equation}
		\label{eqTue}
			\begin{split}
				\mathcal{T}^{ue} = \sqrt{3}\,& \frac{1 + \overline{\gamma}_c\,u}{1 + \overline{\gamma}_c\,u/\sqrt{3}}\left[1 + \frac{\overline{\gamma}_c}
				{2\big(1 + \overline{\gamma}_c\,u/\sqrt{3}\big)^2} \right. \\
				&+\left. \frac{3\overline{\gamma}_c^2}{8\big(1 + \overline{\gamma}_c\,u/\sqrt{3}\big)^2}\right]\,,
			\end{split}
		\end{equation}
		so that it has the appropriate Newtonian limit $\mathcal{T}^{ue}\rightarrow3$.
		This can be checked on the numerical data displayed in Fig.~\ref{FigTroutue}.
		The saturation value in the yielding regime is given by:
		\begin{equation}
		\label{eqTue2}
			\mathcal{T}^{ue}\underset{t_\Gamma \gg t_\gamma}{\longrightarrow}\sqrt{3}\big(1+\overline{\gamma}_c/2+3\overline{\gamma}_c^2/8\big)\,,
		\end{equation}
		which is also inherited from the structure of the effective yield stress of the uniaxial extension flow.
		In particular in that case, $\mathcal{T}^{ue}\geqslant\sqrt{3}\simeq1.73$.

		Another striking difference between $\mathcal{T}^{pe}$ and $\mathcal{T}^{ue}$ in Figs.~\ref{FigTroutpe} and \ref{FigTroutue} is the fact that contrary to the former,
		the latter displays a peak in the Newtonian regime.
		This can be understood from Eqs.~(\ref{eqTpe}) and (\ref{eqTue}) by an expansion around the Newtonian value:
		\begin{equation}
		\label{eqTpeue}
			\begin{split}
				& \mathcal{T}^{pe} \underset{u\gg1}{=} 4 - \frac{4}{\overline{\gamma}_c\,u} + O \left(\frac{1}{u^2}\right)\\
				& \mathcal{T}^{ue} \underset{u\gg1}{=} 3 - \frac{3}{\overline{\gamma}_c\,u}\left[1 + \sqrt{3}\left(\frac{\overline{\gamma}_c}{2} -1\right)\right]
				+ O \left(\frac{1}{u^2}\right)\,.
			\end{split}
		\end{equation}
		In the vicinity of the Newtonian regime, $\mathcal{T}^{pe}$ tends to decrease with Pe, whereas $\mathcal{T}^{ue}$ tends to grow, at least as long as
		$\overline{\gamma}_c\geqslant2(\sqrt{3} -1)/\sqrt{3} \simeq0.845$, which is always realized in practice.
		We can see here again the influence of the presence, in $\sigma_0^{ue}$ of a richer structure than $\sigma_0^{pe}$, notably an additional $\mathcal{K}_1$
		term, which was difficult to detect at the level of $\sigma_0$ itself.

\section{Conclusion}

	In conclusion, this study shows the generalization of the GITT equation to arbitrary incompressible stationary flows.
	This equation has also been used to define quantities such as the viscosity tensor, which allows an easy comparison of different flow geometries,
	and reduce the complexity of the GITT equations to a few integrals dependent on the system's dynamics, and some tensorial structure defined by the flow geometries.
	We showed how, in addition to providing a tool to provide numerical estimates of the rheological observables of granular liquids and suspensions,
	GITT can also allow an analysis of the finer structure of their behavior through the use of analytically solvable toy models that yield simple
	constitutive laws that can be easily compared to numerical simulations and experiments.

	In particular, we have shown that the evolution of $\mu$ in a general flow can be qualitatively quite different from what is expected
	in the simple shear case, with bigger amplitudes of variation and possible nonmonotonous evolution with the shear rate.
	Due to the GITT toy models, we were able to generalize the $\mu(\mathcal{I})$-law to account for these new behaviors; see Eq.~(\ref{eqMuImod}).
	This could be particularly relevant for the study of flows of pastes and granular suspensions, which frequently undergo flow different
	from simple shear in the context of their various industrial applications.
	We hope that our work will help motivate further numerical and experimental studies to investigate these behaviors.

\section*{Acknowledgements}

	This work was funded by the Deutscher Akademischer Austauschdienst (DAAD) and the Deutsche Forschungsgemeinschaft (DFG),
	grant KR 486712.11
	We warmly thank Th. Voigtmann for stimulating discussions and helpful suggestions.
	We thank W.T. Kranz for carefully reading the manuscript.

\appendix

\section{The GITT vertex tensor $\mathbb{J}_{\alpha\beta}^{\quad\omega\theta}$}
\label{AJ}

	In this section, we present the vertex tensor which is obtained after evaluation of the spherical part of the $k$-integral in Eq.~(\ref{eqITT1}):
	\begin{equation}
	\label{AeqdefJ}
		\mathbb{J}_{\alpha\beta}^{\quad\theta\omega} = \int d\Omega \,\frac{1}{2(2\pi)^3}\,\mathcal{V}_{k,\theta\omega}^{\sigma}
		\mathcal{W}_{k,\alpha\beta}^{\sigma}\,,
	\end{equation}
	where $d\Omega$ is the integration over the angular variables.

	For sake of clarity, the following expressions are valid only for a \textit{symmetric} version of the flow tensor $\kappa$ (which excludes the simple
	shear case).
	More general expressions can be worked out without more difficulty in the general case.
	The vertex tensor thus has the following symmetries: $\mathbb{J}_{\alpha\beta}^{\quad\theta\omega}=\mathbb{J}_{\beta\alpha}^{\quad\theta\omega}
	=\mathbb{J}_{\alpha\beta}^{\quad\omega\theta}=\mathbb{J}_{\beta\alpha}^{\quad\omega\theta}$.

	Because of the spherical average, it is convenient to distinguish four typical configurations.

	\subsection{Case $\alpha = \beta$, $\theta = \omega$}

		In that case, according to Eq.~(\ref{eqVW}), the MCT vertices are given by:
		\begin{equation}
			\begin{split}
				& \mathcal{V}_{k,\theta\theta}^{\sigma} = \hat{k}_\theta(-t)\hat{k}_\theta(-t)\, k(-t)\Delta\sigma + \sigma_\perp \\
				& \mathcal{W}_{k,\alpha\alpha}^{\sigma} = \frac{1+\varepsilon}{2\,S_k^2}
				\big(\hat{k}_\alpha\hat{k}_\alpha\, k\,\Delta\sigma + \sigma_\perp\big) \,,
			\end{split}
		\end{equation}
		where $\Delta\sigma$ and $\sigma_\perp$ are given by Eq.~(\ref{eqsDO}).

		Under such conditions, the vertex integral is given by:
		\begin{equation}
		\label{AeqJ1}
			\begin{split}
				&\mathbb{J}_{\alpha\alpha}^{\quad\theta\theta} = \frac{k^2}{k^2(-t)} \frac{1+\varepsilon}{2\,S_k^2} \bigg\{
				\frac{\sigma_\perp^2}{4\pi^2}\frac{k^2(-t)}{k^2} + \frac{\sigma_\perp\Delta\sigma k}{12\pi^2}\frac{k^2(-t)}{k^2}  \\
				& + \frac{k(-t)\Delta\sigma\sigma_\perp}{12\pi^2}\mathbb{P}_\theta
				 +kk(-t)\Delta\sigma^2\Big[\frac{\mathbb{Q}_{\theta\alpha}}{20\pi^2} + \frac{1}{60\pi^2}\sum_{\beta\neq\alpha}
				\mathbb{Q}_{\theta\beta}\Big]\bigg\} \,,
			\end{split}
		\end{equation}
		where we defined:
		\begin{equation}
			\begin{split}
				& \mathbb{P}_\theta = \big(1 + \kappa_{\theta\theta} t\big)^2 + \sum_{\omega\neq\theta}\big(\kappa_{\theta\omega}t\big)^2\\
				& \mathbb{Q}_{\theta\alpha} = \delta_{\theta\alpha} \big(1 + \kappa_{\theta\theta} t\big)^2
				+ \big(1-\delta_{\theta\alpha}\big)\big(\kappa_{\theta\alpha}t\big)^2\,,
			\end{split}
		\end{equation}
		The prefactor $k^2/k^2(-t)$ comes from the definition of the $\hat{\cdot}$ operator.

	\subsection{Case $\alpha = \beta$, $\theta \neq \omega$}

		In that case, according to Eq.~(\ref{eqVW}), the MCT vertices are given by:
		\begin{equation}
			\begin{split}
				& \mathcal{V}_{k,\theta\omega}^{\sigma} = \hat{k}_\theta(-t)\hat{k}_\omega(-t)\, k(-t)\Delta\sigma \\
				& \mathcal{W}_{k,\alpha\alpha}^{\sigma} = \frac{1+\varepsilon}{2\,S_k^2}
				\big(\hat{k}_\alpha\hat{k}_\alpha\, k\,\Delta\sigma + \sigma_\perp\big) \,.
			\end{split}
		\end{equation}

		Under such conditions, the vertex integral is given by:
		\begin{equation}
		\label{AeqJ2}
			\begin{split}
				\mathbb{J}_{\alpha\alpha}^{\quad\theta\theta} &= \frac{k^2}{k^2(-t)} \frac{1+\varepsilon}{2\,S_k^2} \bigg\{
					\frac{k(-t)\Delta\sigma\sigma_\perp}{12\pi^2}\,\sum_\nu \mathbb{N}_{\theta\omega}^{\quad\nu\nu}\\
					&+kk(-t)\Delta\sigma^2\Big[ \frac{\mathbb{N}_{\theta\omega}^{\quad\alpha\alpha}}{20\pi^2}
					+ \sum_{\beta\neq\alpha}\frac{\mathbb{N}_{\theta\omega}^{\quad\beta\beta}}{60\pi^2}\Big]\bigg\} \,,
			\end{split}
		\end{equation}
		where we defined:
		\begin{equation}
			\begin{split}
				\mathbb{N}_{\theta\omega}^{\quad\alpha\alpha} =& \delta_{\alpha\theta}\big(\kappa_{\alpha\omega}t +
				\kappa_{\alpha\alpha}\kappa_{\alpha\omega}t^2 \big) + \delta_{\alpha\omega}\big(\kappa_{\theta\alpha}t +
				\kappa_{\alpha\alpha}\kappa_{\theta\alpha}t^2\big)\\
				& \big(1 - \delta_{\alpha\theta}\big)\big(1 - \delta_{\alpha\omega}\big)\big(\kappa_{\alpha\theta}\kappa_{\alpha\omega}t^2\big)\,.
			\end{split}
		\end{equation}

	\subsection{Case $\alpha \neq \beta$, $\theta = \omega$}

		In that case, according to Eq.~(\ref{eqVW}), the MCT vertices are given by:
		\begin{equation}
			\begin{split}
				& \mathcal{V}_{k,\theta\theta}^{\sigma} = \hat{k}_\theta(-t)\hat{k}_\theta(-t)\, k(-t)\Delta\sigma + \sigma_\perp \\
				& \mathcal{W}_{k,\alpha\beta}^{\sigma} = \frac{1+\varepsilon}{2\,S_k^2}
				\big(\hat{k}_\alpha\hat{k}_\beta\, k\,\Delta\sigma\big) \,.
			\end{split}
		\end{equation}

		Under such conditions, the vertex integral is given by:
		\begin{equation}
		\label{AeqJ3}
			\begin{split}
				\mathbb{J}_{\alpha\alpha}^{\quad\theta\theta} &= \frac{k^2}{k^2(-t)} \frac{1+\varepsilon}{2\,S_k^2}
				\frac{kk(-t)\Delta\sigma^2}{30\pi^2} \mathbb{N}_{\alpha\beta}^{\quad\theta\theta}\,.
			\end{split}
		\end{equation}

	\subsection{Case $\alpha \neq \beta$, $\theta \neq \omega$}

		In that case, according to Eq.~(\ref{eqVW}), the MCT vertices are given by:
		\begin{equation}
			\begin{split}
				& \mathcal{V}_{k,\theta\omega}^{\sigma} = \hat{k}_\theta(-t)\hat{k}_\omega(-t)\, k(-t)\Delta\sigma \\
				& \mathcal{W}_{k,\alpha\beta}^{\sigma} = \frac{1+\varepsilon}{2\,S_k^2}
				\big(\hat{k}_\alpha\hat{k}_\beta\, k\,\Delta\sigma\big) \,.
			\end{split}
		\end{equation}

		Under such conditions, the vertex integral is given by:
		\begin{equation}
		\label{AeqJ4}
			\begin{split}
				\mathbb{J}_{\alpha\alpha}^{\quad\theta\theta} &= \frac{k^2}{k^2(-t)} \frac{1+\varepsilon}{2\,S_k^2}
				\frac{kk(-t)\Delta\sigma^2}{60\pi^2} \mathbb{M}_{\alpha\beta}^{\quad\theta\omega}\,,
			\end{split}
		\end{equation}
		where we defined:
		\begin{equation}
			\begin{split}
				\mathbb{M}_{\alpha\beta}^{\quad\theta\omega} =& \big( \delta_{\alpha\theta}\delta_{\beta\omega} + \delta_{\alpha\omega}\delta_{\beta\theta} \big)
				\Big[1 + \big(\kappa_{\alpha\alpha} + \kappa_{\beta\beta}\big)t \\
				& \quad+ \big(\kappa_{\alpha\alpha}\kappa_{\beta\beta} + \kappa_{\alpha\beta}^2\big) t^2\Big] \\
				& + \Big[ \delta_{\alpha\omega} \big(1-\delta_{\beta\theta}\big) + \delta_{\beta\omega} \big(1-\delta_{\alpha\theta}\big)\Big] \\
				& \quad\times\Big[ \kappa_{\alpha\theta}t + \big( \kappa_{\alpha\beta}\kappa_{\beta\theta} + \kappa_{\beta\beta}\kappa_{\alpha\theta}\big)t^2\Big]\,.
			\end{split}
		\end{equation}

\bibliography{EP.bib}

\end{document}